\documentclass[a4paper]{rsproca_new}
\usepackage{bm}
\usepackage{microtype}
\usepackage{graphicx}
\usepackage{mathrsfs}
\usepackage[usenames,dvipsnames,svgnames,table]{xcolor}

\jname{rspa}

\newcommand{\vect}[1]{\bm{#1}} 
\newcommand{\matr}[1]{#1} 
\newcommand{\morpho}{M} 
\newcommand*{\tran}{{\mkern-1.5mu\mathsf{T}}}

\numberwithin{equation}{section}

\titlehead{Research}

\begin{document}

\title{Simulating shallow morphodynamic flows on evolving topographies}

\author{
Jake Langham$^{1}$, Mark J. Woodhouse$^{2}$, Andrew J. Hogg$^{1}$, Luke T.
Jenkins$^{2}$ and Jeremy C. Phillips$^{2}$}

\address{$^{1}$School of Mathematics, Fry Building, University of Bristol, BS8
1UG, UK\\
$^{2}$School of Earth Sciences, Wills Memorial Building, University of Bristol,
Bristol, BS8 1RJ, UK}

\subject{fluid dynamics, geophysics, numerical modelling}

\keywords{shallow water, morphodynamics, debris flow, well-balanced,
ill posedness, self-accelerating}

\corres{Jake Langham\\
\email{J.Langham@bristol.ac.uk}}

\begin{abstract}
We derive general depth-integrated model equations for overland flows featuring
the evolution of suspended sediment that may be eroded from or deposited
onto the underlying topography (`morphodynamics').  The resulting equations
include geometric corrections that account for large variations in slope
angle.  These are often non-negligible for Earth-surface flows and may
consequently be important for simulating natural hazards.  We also show how
to adapt existing finite volume schemes for the classical shallow water
equations, to simulate our new equations in a way that preserves uniform
steady states and exactly conserves the combined mass of the flow and bed.
Finally, to demonstrate our formulation, we present computations using simple
example model closures, fed by point flux sources.  On initially constant
slopes, flows exhibit different behaviours depending on the grade. Shallow
slopes lead to weakly morphodynamic spreading flows that agree well with
analytical similarity solutions. On more severe slopes, rapid erosion
occurs, leading to self-channelising flows and ultimately a `super-erosive'
state, in which sediment entrainment and gravitational acceleration
perpetually reinforce each other.  
\end{abstract}

\maketitle

\section{Introduction}
Overland flows of water mixed with sediment and debris represent one of the
Earth surface's primary hazards~\cite{Jakob2005,Auker2013,Dowling2014}. 
Such flows enhance their destructive potential by
entraining bed material which is later deposited
downstream~\cite{Pierson1990,Scott2005}.
In many cases, the basal surface can be dramatically reshaped by the flow over
its lifespan.  Since flows are guided by the topography that they propagate
over, there is a mutual coupling between the flow and its bed that becomes
particularly important in highly energetic regimes, where large volumes of
sediment are mobilised by the flow.
Consequently, there is a pressing need for accurate and reliable large-scale
flow models of these systems that incorporate the so-called `morphodynamic'
processes of sediment erosion and deposition.

\begin{figure}
    \begin{centering}%
    \includegraphics[width=\textwidth]{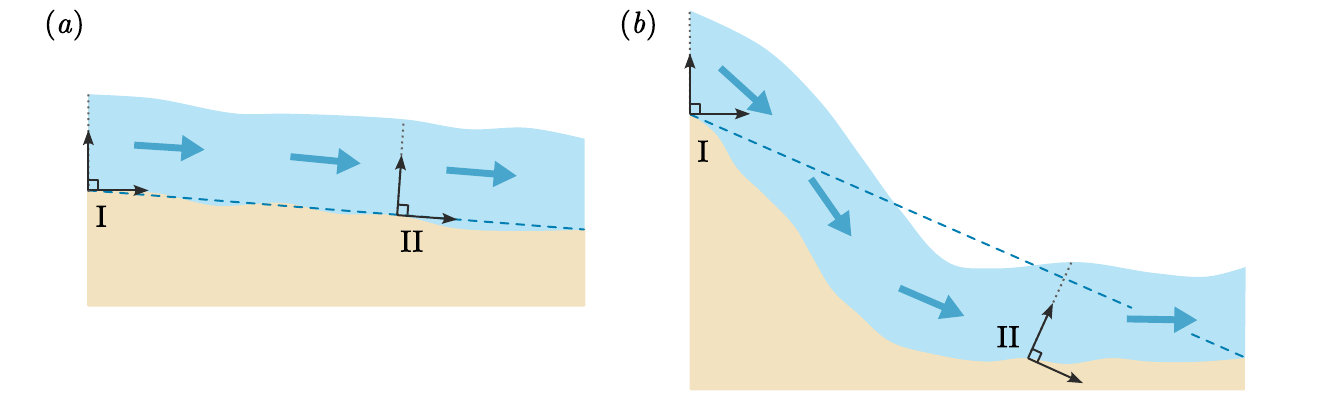}
        \caption{
        Shallow flows featuring (\emph{a})~small and (\emph{b})~large ranges of
        slope angles.  The black arrows labelled show orthogonal coordinate
        frames that are (I) aligned with gravity and (II) the mean slope
        (plotted in dashed blue).  Blue arrows indicate the direction of the
        mean flow velocity. 
        }%
    \label{fig:frames}%
    \end{centering}%
\end{figure}

Many authors have proposed extensions to the classical shallow water equations
in order to account for the particular physics of mixed-sediment morphodynamic
flows~\cite{Capart1998,Brufau2000,Egashira2001,Cao2004,Simpson2006,Wu2007,Zech2008,Armanini2009,Murillo2010,Greco2012,Benkhaldoun2013,Swartenbroekx2013,Li2018,Pudasaini2021}.
These depth-averaged models are a promising step towards achieving truly
predictive debris flow simulations that may be relied upon for hazards
assessment~\cite{Soares2012,Liu2013,Guan2015,Santillan2020,Jenkins2023} and for
estimating physical impacts to the built
environment~\cite{Haugen2008,Gentile2022}.  However, it is now known that many
common morphodynamic formulations are mathematically ill posed as initial value
problems~\cite{Chavarrias2018,Langham2021}, which renders them useless as
potential tools for hazards prediction, since simulations of ill-posed models
cannot be  converged to obtain reliable high resolution reference solutions.
Moreover, as Iverson and Ouyang note in their detailed
review~\cite{Iverson2015}, most prior efforts to produce models have contained
systemic errors in their mathematical derivations.  A particular technical
challenge is to account for geometric effects of the underlying topography in a
way that is consistent with the coordinates used in depth-averaging. 
To see why, consider the sketches in Fig.~\ref{fig:frames}.
Panel~(\emph{a}) depicts flow on a mild slope with relatively small variations
in its basal gradient---a typical configuration encountered in applications
including river dynamics and hydraulic engineering. Two orthonormal frames are
commonly used to derive shallow-layer equations in this setting: (I) gravity
aligned axes and (II) mean slope aligned axes.  Either frame is suitable for
depth-averaging, since in both cases the streamwise axis is roughly parallel to
the mean flow.  Conversely, the right-hand diagram
[Fig.~\ref{fig:frames}(\emph{b})] portrays the situation for many debris flows,
which are initiated on steep slopes before running out onto flatter terrain.
Averaging the governing equations with respect to either frame will introduce
systematic errors in locations where the mean flow and coordinate axes are
misaligned.

Though
multiple frameworks exist which resolve this issue in classical shallow water
models~\cite{Dressler1978,Berger1998,Gray1999,Keller2003,Bouchut2003,Bouchut2004,Peruzzetto2021},
these have largely not been generalised to include the effect of basal
topography that is morphodynamically coupled to the flow.  An exception
is Bouchut \emph{et al.}~\cite{Bouchut2008}, who have developed shallow
morphodynamic avalanche models in one spatial dimension, for both arbitrary
basal topographies and reduced cases where slope angle variation and bed
dynamics are the same order as the flow depth.

Even in the non-morphodynamic case, models that include these geometric effects
appear to have so far made little to no impact on operational debris flow
modelling in mountainous regions in the past few decades, despite the fact that
these flows necessarily propagate over undulating terrains with large variations
in slope angle.  This is likely in part due to the additional complexity that
the governing equations must feature, as well as the need to specialise models
for particular settings.  The recent work of Peruzzetto \emph{et
al.}~\cite{Peruzzetto2021} has made significant progress in both of these
regards, by deriving a non-morphodynamic shallow debris flow model with
curvature terms and showing that it is feasible to numerically solve it in
settings relevant for hazards simulations.

The work in this paper divides into three parts.  Firstly, in
Sec.~\ref{sec:derivation} we develop well-posed depth-averaged morphodynamic
model equations for debris flows in two spatial dimensions in a way that
accounts for large and time-dependent variations in topographic slope angle.
While part of our aim is to progress the theoretical advances of previous
studies, we place primary focus on obtaining and implementing a framework that
is practical for operational use.  To this end, restrictions are imposed on both
the curvature and dynamics of the bed to obtain a generic and relatively
straightforward system that may be specialised to particular flows through the
selection of closures specifying the rheology and morphodynamics.

Secondly, in Sec.~\ref{sec:numerics}, we show how to solve our equations
numerically, by adapting existing finite volume shallow water schemes.  This is
made convenient by use of an operator splitting approach, which separates the
hydraulic evolution of the flow and the morphodynamics, so that relatively few
adjustments need to be made to the flow solver itself.
Although many studies have already described schemes for the current crop of
shallow morphodynamic
models~\cite{Brufau2000,Cao2004,Simpson2006,Wu2007,Yue2008,Armanini2009,Murillo2010,Xia2010,Greco2012,Benkhaldoun2013,Swartenbroekx2013,Canelas2013,Liu2015,Liu2017,Li2018,Martinez2019,Chertock2020},
we identify and overcome specific issues that must be accounted for when these
models are extended to include more geometrical information.  Particular care is
taken to obtain a scheme that exactly conserves mass---a consideration which is
somewhat subtle in our framework because local volume elements depend on the
slope angle.  When time-stepping the morphodynamics, this amounts to ensuring
that any change in the bed volume corresponds exactly with material added to or
deposited from the flow.  An auxiliary correction algorithm preserves the
positivity of the flow depth and solids concentration in cases of rapid
sediment deposition.  Furthermore, we adapt the approach of
Kurganov and Petrova~\cite{Kurganov2007}, to exactly compute steady states, including stationary
solutions (so-called `lake-at-rest' states). Such schemes are often referred to
as `well balanced'. Numerical solvers which fail
to compute these states accurately generate spurious momentum from static
initial conditions and may even begin to erode the topography as a result.

Finally, to demonstrate the success of our approach, in Sec.~\ref{sec:results}
we present some illustrative results. Ideally, numerical methods should be
tested by comparing their outputs with analytical reference solutions. However,
closed-form solutions are difficult to obtain in most circumstances when
morphodynamics is present. Nevertheless, for flows from point sources on planes
of initially constant gradient that feature only weak erosion, we
derive similarity solutions and confirm that our numerical results agree with
them when slopes are mild.  On steeper slopes, erosion becomes more rapid and
can cause flows to self-channelise. When slopes are extreme, flows ultimately
enter a regime where the mutual coupling of sediment entrainment and forcing
from the weight of the mixture forms a self-perpetuating cycle that accelerates
the flow.  We also briefly demonstrate flow on a surface that features a slope
transition between two different gradients and note that its behaviour can be
largely understood by combining insights from the milder and steeper
constant-slope regimes.

\section{Derivation of the governing equations}
\label{sec:derivation}%
We consider a well-mixed flow of fluid and solid particles,
propagating over an erodible substrate, featuring mass transfer between the flow and
its underlying bed.  The particles are assumed to be monodisperse and small,
relative to the flow length scales, so that both fluid and solid phases may be
modelled as continua.  Moreover, we suppose that the substrate is static and
homogeneously composed of the same materials as the flow, with the solid
particles occupying a constant fraction $\psi_b$ of its volume.  The densities
of fluid and solids are denoted by~$\rho_f$ and~$\rho_s$ respectively.  Then,
the total density field for the mixture is $\rho = \psi\rho_s + (1-\psi)\rho_f$,
where $\psi$ is the scalar field quantifying the volume fraction of solids in
the flow.  
A velocity field for the mixture may be defined by summing
the contributions of each fraction to the bulk momentum:
\begin{equation}
    \psi\rho_s\vect{u}_s 
    +
    (1-\psi)\rho_f\vect{u}_f
    \equiv \rho\vect{u},
    \label{eq:total mom}%
\end{equation}
where $\vect{u}_s$ and $\vect{u}_f$ denote the velocities of the solid and fluid
phases respectively~\cite{Iverson2015}.

The phases obey separate conservation laws for their occupying volume and
momentum, which may be linearly combined as necessary to obtain governing
equations for the mixture.  In terms of Cartesian coordinates $(x,y,z)$
(oriented with gravity pointing in the negative $z$-direction) and time $t$,
we write conservation of bulk mass, solids volume and bulk momentum, as
\begin{subequations}
\begin{gather}
    \frac{\partial \rho}{\partial t} + 
    \nabla\cdot\left(\rho\vect{u}\right) = 0,\label{eq:continuity 1}\\
    \frac{\partial \psi}{\partial t}
    + \nabla\cdot(\psi\vect{u}_s) = 0,\label{eq:continuity 2}\\
    \rho\left[\frac{\partial \vect{u}}{\partial t}
    +
    \vect{u}\cdot\nabla\vect{u}\right]
    + \nabla \cdot \kappa
    = -\nabla p + \nabla \cdot \matr{\tau} - \rho\vect{g},
    \label{eq:cart mom}%
\end{gather}
    \label{eq:3d eqs}%
\end{subequations}
where $\nabla \equiv (\partial /\partial x, \partial / \partial y, \partial /
\partial z)^\tran$\!.  On the left-hand side of Eq.~\eqref{eq:cart mom},
$\nabla\cdot\kappa$ is momentum transport by the interphase drift velocity
$\vect{u}_s-\vect{u}_f$, with
$\kappa=\rho_s\rho_f\psi(1-\psi)\rho^{-1}(\vect{u}_s -
\vect{u}_f)\otimes(\vect{u}_s-\vect{u}_f)$.  On the right-hand side, $p$ denotes
the pressure field of the mixture, $\matr{\tau}$ denotes its deviatoric stress
tensor and $\vect{g}=(0,0,g)^\tran$ is gravitational acceleration.  We have not
included the separate single-phase momentum equations here, since these will not
ultimately be required to close our model.

\begin{figure}
    \begin{centering}%
    \includegraphics[width=\textwidth]{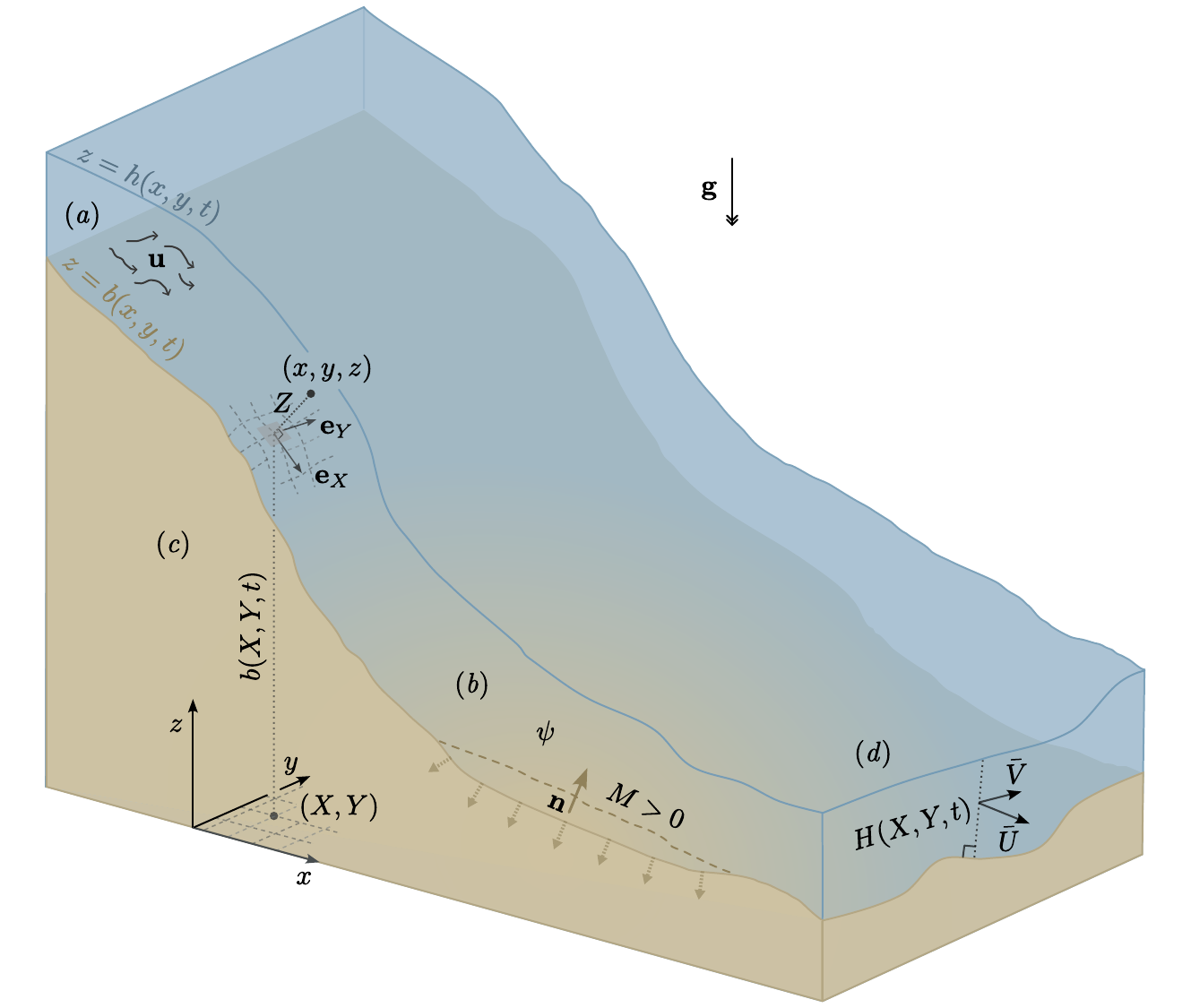}
        \caption{Diagram of the flow system, 
        depicting aspects of the model setup and derivation.  (\emph{a})~Flow
        bounding surfaces and the velocity field $\vect{u}$, which is
        predominantly oriented parallel to the bed. (\emph{b})~Depiction
        of morphodynamics. Erosion at rate $M > 0$ extracts a patch material
        from the bed and adds it to the flow, locally enhancing the solids
        fraction field $\psi$ and leading to a preferentially concentrated
        region, as indicated by the surrounding shading of the flow.
        (\emph{c})~Relationship between Cartesian $(x,y,z)$ and surface-fitted
        coordinates $(X,Y,Z)$.  (\emph{d})~Surface-fitted velocity fields
        $\bar{U}$ and $\bar{V}$, obtained after averaging over the depth $H(X,
        Y, t)$, measured along the direction normal to the bed.
        }%
    \label{fig:schematic}%
    \end{centering}%
\end{figure}

For reference, a schematic of the flow, its geometry, and key variables used in
the following derivation, is depicted in Fig.~\ref{fig:schematic}.
As highlighted in Fig.~\ref{fig:schematic}(\emph{a}), the flowing mixture is
bounded below by the bed substrate, whose boundary is given by the surface
$F_b(x,y,z,t) \equiv z - b(x,y,t) = 0$.  As depicted in
Fig.~\ref{fig:schematic}(\emph{b}), this boundary evolves according to
morphodynamic processes, which are assumed to act normal to the bed surface, at
a rate $M(x,y,t)$. More precisely, we stipulate that $F_b$ moves according to
\begin{equation}
    \frac{\partial F_b}{\partial t} - M\vect{n}\cdot \nabla F_b =
    0,
    \label{eq:bc bed}%
\end{equation}
where $\vect{n}$ is the outward-pointing unit vector field normal to the bed
surface.
The sign convention of Eq.~\eqref{eq:bc bed} is that $M(x,y,t) > 0$ implies net
erosion and $M(x,y,t) < 0$ implies net deposition.
By definition, $\vect{n} = \nabla F_b / |\nabla F_b|$ and so
\begin{equation}
    \vect{n} = \frac{1}{\gamma}\left(-\frac{\partial b}{\partial x}, -\frac{\partial
    b}{\partial y}, 1\right)^{\!\tran},
    \quad \textrm{where} \quad
    \gamma = \sqrt{1 + \left(\frac{\partial b}{\partial x}\right)^{\!\!2} + 
    \left(\frac{\partial b}{\partial y}\right)^{\!\!2}}.
    \label{eq:n}%
\end{equation}
Note that $\gamma = 1/\cos(\vartheta)$, where $\vartheta$ is the angle between the
local slope normal and gravity. 
We may then write the equation
\begin{equation}
    \frac{\partial b}{\partial t} = -\gamma M,
    \label{eq:bed evolution}%
\end{equation}
for the bed evolution, leaving the morphodynamic rate function $M$ as a general
closure that must be specified for particular systems.

Across the bed-flow interface, there are discontinuous jumps in the physical
quantities from their assumed basal values, to the flowing state. We label
variables evaluated on the immediate flow side of the discontinuity with ${}^+$
superscripts. Since mass and volume must be conserved over the jump, any motion of the
interface must be balanced by corresponding fluxes on the flowing side.  From
Eqs.~\eqref{eq:continuity 1} and~\eqref{eq:continuity 2}, we determine that
\begin{subequations}
\begin{gather}
    \rho^+ \vect{u}^+ \cdot \vect{n} = (\rho_b - \rho^+) M,
    \quad\text{and}\quad
    \psi^+ \vect{u}_s^+ \cdot \vect{n} = (\psi_b - \psi^+) M,
    \tag{\theequation\emph{a,b}}%
    \label{eq:cart jump bcs}%
\end{gather}
\end{subequations}
where $\rho_b = \psi_b\rho_s + (1-\psi_b)\rho_f$ is the density of the bed.

The upper boundary of the flow is a free surface with equation $F_h(x,y,z,t) =
h(x,y,t) - z = 0$.  Since there is no mass transfer across this interface, its
boundary is kinematic and evolves according to
\begin{equation}
\frac{\partial F_h}{\partial t} + \vect{u}|_h \cdot \nabla F_h = 0,
    \label{eq:bc surface}
\end{equation}
where $\vect{u}|_h=\vect{u}|_{z=h(x,y,t)}\equiv (u_h, v_h, w_h)^\tran$.
Mechanically, the mixture is stress-free and at atmospheric pressure across this
surface.
Conservation of fluid and solids volume necessitates $\vect{u}_s =
\vect{u}_f$ along the flow side of the interface, so this is the only 
condition required at this boundary.

A shallow-layer model for the flow may be obtained by depth-averaging
Eqs.~(\ref{eq:3d eqs}\emph{a--c}), subject to the boundary
conditions~\eqref{eq:cart jump bcs} and~\eqref{eq:bc surface}, and systematically
neglecting fluid accelerations that are perpendicular to the plane of motion.
In cases where the bed possesses steep gradients, this means that the
leading-order dynamics retains non-negligible vertical
acceleration~\cite{Iverson2015}.
Moreover, if the topography contains a mixture of steep and
shallow grades it is not possible to fix a Cartesian frame such that the 
motion may be consistently neglected in a particular coordinate direction.
Therefore, in order to obtain a model that may be applied for general
topographies, we transform to a curvilinear frame that is fitted to the
time-dependent bed surface.

Specifically, we rewrite each point in terms of new time-dependent coordinates
$(X,Y,Z)$, as
\begin{equation}
    \begin{pmatrix}x \\ y \\ z
    \end{pmatrix}
    =
    \begin{pmatrix}
        X \\ Y \\ b(X, Y, t)
    \end{pmatrix}
    +
    Z\vect{n}(X,Y,t).
    \label{eq:xyz}%
\end{equation}
Note that at each point in time, $Z=0$ represents a Cartesian parametrisation of
the bed surface and the coordinate $Z \geq 0$ gives the perpendicular distance
from the bed.  The mapping is given diagrammatically in
Fig.~\ref{fig:schematic}(\emph{c}).  It is unique, provided that the local
radius of curvature everywhere exceeds the flow depth and that $\vect{n}$ is
everywhere well defined and single valued.
Variants and generalisations of this
transformation have previously been employed to develop non-morphodynamic
shallow-layer models~\cite{Keller2003,Bouchut2003,Bouchut2004,Peruzzetto2021}.
Here, we extend this approach for flows with a time-dependent bed.

The basis vectors $\{\vect{e}_X,\vect{e}_Y,\vect{e}_Z\}$ for the new frame are
obtained by differentiating Eq.~\eqref{eq:xyz}:
\begin{subequations}
\begin{equation}
    \vect{e}_X = \begin{pmatrix}
        1\\ 0\\ \frac{\partial b}{\partial X}
    \end{pmatrix}+ Z \frac{\partial \vect{n}}{\partial X},
    \quad
    \vect{e}_Y = \begin{pmatrix}
        0\\ 1\\ \frac{\partial b}{\partial Y}
    \end{pmatrix}+ Z \frac{\partial \vect{n}}{\partial Y},
    \quad
    \vect{e}_Z = \vect{n}.
    \tag{\theequation\emph{a--c}}%
\end{equation}
\label{eq:basis vecs}%
\end{subequations}
These are time dependent and neither mutually orthogonal, nor unit length. We
define $\matr{A}$ to be the change of basis matrix from Cartesian to the new
frame and denote $J = \det(\matr{A}^{-1})$. We shall write the bulk flow
velocity in slope-aligned coordinates as $\vect{U} = (U,V,W)^\tran =
\matr{A}\vect{u}$ and the solids velocity as $\vect{U}_s =
(U_s,V_s,W_s)^\tran=\matr{A}\vect{u}_s$.  Similarly, the interphase momentum
transport and stress tensors are to be denoted as $\matr{K} = \matr{A}
\matr{\kappa} \matr{A}^{-1}$ and $\matr{T} = \matr{A} \matr{\tau} \matr{A}^{-1}$
respectively.  Time derivatives in the new coordinates must be transformed to
account for the advection of the reference frame with velocity $\dot{\vect{X}}
\equiv (\partial X/\partial t, \partial Y/\partial t, \partial Z/\partial
t)^\tran$.  By taking the total derivative of Eq.~\eqref{eq:xyz} with respect to
time (i.e.\ holding the Cartesian coordinates fixed) and rearranging, we compute
\begin{equation}
    \dot{\vect{X}} = 
    -\frac{1}{\gamma}\frac{\partial b}{\partial t} 
    \!\begin{pmatrix}
        \frac{\partial b}{\partial X}/\gamma\\
        \frac{\partial b}{\partial Y}/\gamma\\
    1\end{pmatrix}
    +
    Z
    \frac{\mathrm{d}~}{\mathrm{d} t}
    \!\begin{pmatrix}
        \frac{\partial b}{\partial X}/\gamma\\
        \frac{\partial b}{\partial Y}/\gamma\\
        0
    \end{pmatrix}\!.
    \label{eq:dXdt}%
\end{equation}
Spatial derivatives may be transformed using the chain rule, which dictates that
$\nabla = A^\tran \nabla_{\vect{X}}$, where
$\nabla_{\vect{X}} \equiv (\partial/\partial X,\partial/\partial Y,
\partial/\partial Z)^\tran$ is the slope-aligned gradient operator.

The governing equations written with respect to this frame, may then be determined to
be
\begin{subequations}
\begin{gather}
    \frac{\partial \rho}{\partial t}
    + \dot{\vect{X}}\cdot\nabla_{\vect{X}}\rho
    + \frac{1}{J}\nabla_{\vect{X}}\cdot(J\rho\vect{U}) = 0,\\
    \frac{\partial \psi}{\partial t}
    + \dot{\vect{X}}\cdot\nabla_{\vect{X}}\psi
    + \frac{1}{J}\nabla_{\vect{X}}\cdot(J\psi\vect{U}_s) = 0,\\
    \rho\left[\frac{\partial\vect{U}}{\partial t}
    + \dot{\vect{X}}\cdot\nabla_{\vect{X}}\vect{U}
    + \vect{U}\cdot\nabla_{\vect{X}}\vect{U}
    \right]
    + \matr{A}\matr{A}^\tran\nabla_{\vect{X}}p - \vect{S}
    - \rho\matr{A}\vect{g} = \vect{R},
\end{gather}
    \label{eq:3d governing eqs}%
\end{subequations}
where $\vect{S}$ the transformed divergence of the combined interphase momentum
transport and stress tensors, and $\vect{R}$ is a vector containing any terms
that depend on time and space derivatives of the coordinate transformation
matrices. To specify the components of $\vect{S}$ and $\vect{R}$, we adopt an
alternative labelling of the spatial directions $(X_1,X_2,X_3) \equiv (X,Y,Z)$
and employ the Einstein summation convention hereafter, writing 
\begin{subequations}
\begin{gather}
    S_i = A_{jk}A_{lk}\frac{\partial (\matr{T}-\matr{K})_{il}}{\partial X_j},\\
    R_i = -\rho \matr{A}_{ij} \left[
        \frac{\partial \matr{A}^{-1}_{jk}}{\partial t}
        + (\dot{X}_l+ U_l) \frac{\partial \matr{A}^{-1}_{jk}}{\partial
        X_l}
        \right]U_k
    + \matr{A}_{ij} \matr{A}_{lk} (\matr{T}-\matr{K})_{mn} \frac{\partial~}{\partial X_l}
    \left(
    \matr{A}_{jm}^{-1}\matr{A}_{nk}
    \right)\!.
\end{gather}
\end{subequations}
The boundary conditions at the bed transform straightforwardly to 
\begin{subequations}
\begin{equation}
    \rho^+W^+ = (\rho_b - \rho^+) M
    \quad\textrm{and}\quad
    \psi^+W_s^+ = (\psi_b - \psi^+) M.
    \tag{\theequation\emph{a,b}}%
    \label{eq:normal bed bcs}%
\end{equation}
\end{subequations}
At the free surface, we note that $F_h(x,y,z,t) = (H - Z)/\gamma$, where
$H\equiv H(X(t),Y(t),t)$ is a reparametrisation of $h$ in the surface-fitted
coordinates. Therefore, we may transform Eq.~\eqref{eq:bc surface} and evaluate the
resulting equation across the interface, to give
\begin{equation}
    \frac{\partial H}{\partial t} + 
    \left( \frac{\partial X}{\partial t} + U_H \right)
    \frac{\partial H}{\partial X} +
    \left( \frac{\partial Y}{\partial t} + V_H \right)
    \frac{\partial H}{\partial Y} 
    - W_H
    = \frac{\partial Z}{\partial t},
    \quad\textrm{at}~Z=H,
    \label{eq:dHdt}%
\end{equation}
where $\vect{U}|_{Z=H(X,Y,t)}\equiv (U_H, V_H, W_H)^\tran$.
 
To obtain a shallow-layer model, we consider flow regimes in which the
characteristic thickness of the flow $\mathscr{H}$ is far smaller than its
characteristic streamwise length scale $\mathscr{L}$.  This suggests the
ordering parameter $\epsilon = \mathscr{H}/\mathscr{L} \ll 1$.  The topography
has its own scale.  We permit basal gradients of any size, but choose to
constrain the topography by insisting that its characteristic radius of
curvature is nowhere smaller than $O(\epsilon^{-1})$.  Furthermore, we assume
that the rate of change in the bed height due to morphodynamics is no greater
than the normal velocity scale.  These considerations motivate the following
variable rescalings:
\begin{subequations}
\begin{gather}
    Z \mapsto \epsilon Z,~~
    t \mapsto \epsilon^{-\alpha} t,~~
    U \mapsto \epsilon^\alpha U,~~
    V \mapsto \epsilon^\alpha V,~~
    W \mapsto \epsilon^{\alpha+1} W,~~
    p \mapsto \epsilon p,~~
    T_{ij} \mapsto \epsilon^{2\alpha+1} T_{ij},\tag{\theequation\emph{a--g}}\\
    M \mapsto \epsilon^{\alpha+1}M,\quad
    \frac{\partial b}{\partial t} \mapsto \epsilon^{\alpha+1}\frac{\partial
    b}{\partial t},\quad
    \frac{\partial^2 b}{\partial t\partial X_k} \mapsto
    \epsilon^{\alpha+1}\frac{\partial^2
    b}{\partial t \partial X_k}, \quad
    \frac{\partial^2 b}{\partial X_k X_l} \mapsto
    \epsilon\frac{\partial^2 b}{\partial X_k X_l},\tag{\theequation\emph{h--k}}
\end{gather}
for all $i,j\in\{1,2,3\}$ and $k,l\in\{1,2\}$.  We have included a free
parameter $\alpha \geq 0$ in these transformations.
The choice to rescale time with respect to this
scale will later provide a concrete means of justifying the inclusion of
hydrostatic pressure gradients, which becomes important when $\alpha$
exceeds zero. We further assume at this stage, that the velocity difference
between the fluid and solid phases is negligible along the plane parallel to
the bed slope (so $U = U_s$ and $V = V_s$, to leading order).  
The corresponding velocities in
the normal direction must also scale in accordance with the shallowness
assumption.  Therefore, we make the additional rescalings
\begin{equation}
    W_s \mapsto \epsilon^{\alpha + 1} W_s
    \quad \textrm{and} \quad
    K_{ij} \mapsto \epsilon^{2\alpha + 2} K_{ij},
    \tag{\theequation\emph{l,m}}%
\end{equation}
\end{subequations}
for all $i,j\in\{1,2,3\}$.

The above transformations may be substituted as necessary into Eqs.~(\ref{eq:basis
vecs}\emph{a--c}) and~\eqref{eq:dXdt} to determine the leading order components
of the coordinate change. In particular, we compute $J = \gamma +
O(\epsilon^{2})$ and
\begin{subequations}
\begin{equation}
    G \equiv AA^\tran = 
    \begin{pmatrix}
        1 - \!\frac{1}{\gamma^2}\left[\frac{\partial b}{\partial X}\right]^{2}
        &
        -\frac{1}{\gamma^2}\frac{\partial b}{\partial X}\frac{\partial b}{\partial Y} 
        &
        0 \\
        -\frac{1}{\gamma^2}\frac{\partial b}{\partial X}\frac{\partial b}{\partial Y} 
        &
        1 - \!\frac{1}{\gamma^2}\left[\frac{\partial b}{\partial Y}\right]^{2}
        &
        0\\
        0 & 0 & 1
    \end{pmatrix}
    + O(\epsilon^{2}),
    ~
    \dot{\vect{X}}=
    -\frac{\epsilon^{\alpha+1}}{\gamma^2}\frac{\partial b}{\partial t}
    \begin{pmatrix}
        \frac{\partial b}{\partial X}\\
        \frac{\partial b}{\partial Y}\\
        \gamma
    \end{pmatrix}
    +O(\epsilon^{\alpha+2}).
    \tag{\theequation\emph{a,b}}%
\end{equation}
\end{subequations}
Using these expressions, we proceed to apply our variable rescalings to the
governing equations~(\ref{eq:3d governing eqs}\emph{a}--\emph{c}).
After simplifying and neglecting any subdominant terms, we obtain 
\begin{subequations}
\begin{gather}
    \frac{\partial \rho}{\partial t}
    + M\frac{\partial \rho}{\partial Z}
    + \frac{\partial~}{\partial X}(\rho U)
    + \frac{\partial~}{\partial Y}(\rho V)
    + \frac{\partial~}{\partial Z}(\rho W)
    =0,
    \label{eq:rescaled eqs 1}\\
    \frac{\partial \psi}{\partial t}
    + M\frac{\partial \psi}{\partial Z} + 
    \frac{\partial~}{\partial X}(\psi U) +  
    \frac{\partial~}{\partial Y}(\psi V) +  
    \frac{\partial~}{\partial Z}(\psi W_s)
    = 0,
    \label{eq:rescaled eqs 2}\\
    \rho\left[
    \frac{\partial U}{\partial t}
    +M\frac{\partial U}{\partial Z}
    +\vect{U}\cdot\nabla_{\vect{X}}U
    \right]
    = 
    -\epsilon^{-2\alpha} \frac{\rho g}{\gamma^2}\frac{\partial b}{\partial X}
    -\epsilon^{1-2\alpha}\left(G_{11}\frac{\partial p}{\partial X}
    +G_{12}\frac{\partial p}{\partial Y}\right)
    +\frac{\partial T_{13}}{\partial Z},
    \label{eq:rescaled eqs 3}\\
    \rho\left[
    \frac{\partial V}{\partial t}
    +M\frac{\partial V}{\partial Z}
    +\vect{U}\cdot\nabla_{\vect{X}}V
    \right]
    = 
    -\epsilon^{-2\alpha} \frac{\rho g}{\gamma^2}\frac{\partial b}{\partial Y}
    -\epsilon^{1-2\alpha}\left(G_{21}\frac{\partial p}{\partial X}
    +G_{22}\frac{\partial p}{\partial Y}\right)
    +\frac{\partial T_{23}}{\partial Z},
    \label{eq:rescaled eqs 4}\\
    \frac{\partial p}{\partial Z} 
    = -\frac{\rho g}{\gamma} + \epsilon^{2\alpha}\frac{\partial T_{33}}{\partial
    Z},
    \label{eq:rescaled eqs 5}%
\end{gather}
\label{eq:rescaled eqs}%
\end{subequations}
where $G_{ij}$ are the components of the matrix $G = AA^\tran$.
The corresponding basal boundary conditions [Eqs.~\eqref{eq:normal bed bcs}] are
unmodified by the variable transformations, while the constraint on the free
surface becomes
\begin{gather}
    \frac{\partial H}{\partial t} + U_H \frac{\partial H}{\partial X}
    + V_H \frac{\partial H}{\partial Y} - W_H
    = M.\label{eq:rescaled h bc}
\end{gather}
We note that these reduced equations contain neither curvature terms nor time
derivatives of the basal gradients, since these only appear at higher order.
However, these could be reintroduced systematically to obtain a model that
includes these effects, as in Refs.~\cite{Bouchut2004,Bouchut2008,Peruzzetto2021}.

Equations~(\ref{eq:rescaled eqs}\emph{a--e}) are to be integrated with respect to the flow depth measured
perpendicular to the bed, to obtain
depth-averaged flow in two spatial dimensions
as depicted earlier, in Fig.~\ref{fig:schematic}(\emph{d}).
First, we define the quantities 
\begin{subequations}
\begin{gather}
    \bar\rho = \frac{1}{H}\int_0^H \rho \,\mathrm{d}Z,\quad
    \bar\psi = \frac{1}{H}\int_0^H \psi \,\mathrm{d}Z,\quad
    \bar U = \frac{1}{H}\int_0^H U \,\mathrm{d}Z,\quad
    \bar V = \frac{1}{H}\int_0^H V \,\mathrm{d}Z.
    \tag{\theequation\emph{a--d}}%
\end{gather}
\end{subequations}
Note that $\bar\rho$ and $\bar\psi$ are related, via
\begin{equation}
    \bar{\rho} = \rho_f + (\rho_s - \rho_f)\bar\psi.\label{eq:rho}%
\end{equation}
Then, integrating Eqs.~\eqref{eq:rescaled eqs 1},~\eqref{eq:rescaled eqs 2} and
simplifying using the Leibniz rule and the boundary 
conditions in Eqs.~\eqref{eq:normal bed bcs} and~\eqref{eq:rescaled h bc}, gives
\begin{subequations}
\begin{gather}
    \frac{\partial ~}{\partial t}(\bar\rho H)
    +
    \frac{\partial~}{\partial X}(\bar\rho H \bar U)
    +
    \frac{\partial~}{\partial Y}(\bar\rho H \bar V)
    = 
    \rho_b\morpho,\label{eq:dt rho h}\\
    \frac{\partial~}{\partial t}(\bar\psi H)
    +
    \frac{\partial~}{\partial X}(\bar\psi H \bar U)
    +
    \frac{\partial~}{\partial Y}(\bar\psi H \bar V)
    = 
    \psi_b M.\label{eq:dt psi h}%
\end{gather}
\label{eq:shallow mass eqs}%
\end{subequations}
We have implicitly simplified these equations by
omitting terms that arise when averaging over products of the flow
variables. 
For example, in full generality, the second term in Eq.~\eqref{eq:dt rho h}
is
\begin{equation}
    \frac{\partial~}{\partial X}
    \left[
        \bar\rho H \bar U
        +
        \int_0^H (\rho - \bar\rho) (U - \bar U)\,\mathrm{d}Z
    \right]\!.
\label{eq:shape factor}%
\end{equation}
However, the common assumption that any slope-normal shear in the velocity
profile is negligible permits the simplification of this term and its
$Y$-directed counterpart~\cite{Savage1989,Iverson1997,Hogg2004}.  A similar
assumption is used to derive the advective terms in Eq.~\eqref{eq:dt psi h}.

Equation~\eqref{eq:rescaled eqs 5} may be integrated to obtain hydrostatic
pressure, plus a contribution from normal material stresses:
\begin{equation}
    p(Z) = \frac{\bar\rho g(H - Z)}{\gamma}+
    \epsilon^{2\alpha}[T_{33}(Z) - T_{33}(H)].
\end{equation}
Then we compute
\begin{equation}
    \int^{H}_0 \frac{\partial p}{\partial X}\,\mathrm{d}Z
    =
    \frac{\partial~}{\partial X} 
    \left(
    \frac{\bar\rho g H^2}{2\gamma}
    \right)
    +
    \epsilon^{2\alpha} \int^H_0 \frac{\partial
    T_{33}}{\partial X}\,\mathrm{d}Z.
    \label{eq:int dpdX}%
\end{equation}
A similar formula may be obtained for the depth integral of $\partial p /
\partial Y$.  We now integrate the momentum equations~\eqref{eq:rescaled eqs 3}
and~\eqref{eq:rescaled eqs 4} over the flow depth, making use of the boundary
conditions~\eqref{eq:normal bed bcs},~\eqref{eq:rescaled h bc} and simplifying
the pressure gradient terms using Eq.~\eqref{eq:int dpdX} and its $Y$-directed
counterpart, ultimately obtaining
\begin{subequations}
\begin{gather}
    \begin{split}
    \frac{\partial~}{\partial t}(\bar\rho H \bar U)
    +
    \frac{\partial~}{\partial X}(\bar\rho H \bar U^2)
    +
    \frac{\partial~}{\partial Y}(\bar\rho H \bar U \bar V)
    =
    -\epsilon^{-2\alpha}\frac{\bar\rho g H}{\gamma^2}\frac{\partial b}{\partial X}
    \qquad\qquad 
    \qquad\qquad 
    \qquad\qquad \\
    \quad\quad-\epsilon^{1-2\alpha}
    \left(
    G_{11}
    \frac{\partial~}{\partial X}
    + G_{12}
    \frac{\partial~}{\partial Y}
    \right)\!
    \left(
    \frac{\bar\rho g H^2}{2 \gamma}
    \right)
    -T_{13}|_{Z=0^+}
    +
     \rho_b U^+ \morpho
    + \epsilon\int^H_0 \frac{\partial T_{33}}{\partial
    X}\,\mathrm{d}Z,\label{eq:mom w O(eps) 1}\end{split}\\
    \begin{split}
    \frac{\partial~}{\partial t}(\bar\rho H \bar V)
    +
    \frac{\partial~}{\partial X}(\bar\rho H \bar U\bar V)
    +
    \frac{\partial~}{\partial Y}(\bar\rho H \bar V^2)
    =
    -\epsilon^{-2\alpha}\frac{\bar\rho g H}{\gamma^2}\frac{\partial b}{\partial Y}
    \qquad\qquad 
    \qquad\qquad 
    \qquad\qquad \\
    \quad\quad-\epsilon^{1-2\alpha}
    \left(
    G_{21}
    \frac{\partial~}{\partial X}
    + G_{22}
    \frac{\partial~}{\partial Y}
    \right)\!
    \left(
    \frac{\bar\rho g H^2}{2 \gamma}
    \right)
    -T_{23}|_{Z=0^+}
    +
     \rho_b V^+ \morpho
    + \epsilon\int^H_0 \frac{\partial T_{33}}{\partial
    Y}\,\mathrm{d}Z,
        \label{eq:mom w O(eps) 2}%
    \end{split}%
\end{gather}
\label{eq:mom w O(eps)}%
\end{subequations}
Note that terms analogous to Eq.~\eqref{eq:shape
factor} have been simplified by assuming negligible velocity shear.

The role of the right-hand side forcing terms in Eqs.~\eqref{eq:mom w O(eps) 1}
and~\eqref{eq:mom w O(eps) 2} depends on the time scale set by $\alpha$. If
$\alpha = 0$ (so that time is not rescaled) then the hydrostatic pressure
gradient is subdominant. This occurs when the basal gradients are $O(1)$ and the
dynamics are predominantly forced by gravitational acceleration. If instead the
gradients are $O(\epsilon)$, then $\alpha = 1$ and the hydrostatic pressure
gradient is the same order as gravitational forcing.  In order to obtain a model
which is appropriate for both steep and gentle topographies, we retain all
expressions on the right-hand side which are leading order for any $\alpha \geq
0$. The final terms on the right-hand sides of Eqs.~(\ref{eq:mom w
O(eps)}\emph{a,b}) are $O(\epsilon)$ and are not retained henceforth.
Furthermore, we assume at this stage the the terms containing the basal slip
velocities $U^+$ and $V^+$ are comparatively small and make implicit the
notation that the stress components are to be evaluated at the upper bed
surface. On returning to the un-rescaled variables, this leads to the following
expressions of momentum balance
\begin{subequations}
\begin{gather}
    \begin{split}%
    \frac{\partial~}{\partial t}(\bar\rho H \bar U)
    +
    \frac{\partial~}{\partial X}\!\left(\bar\rho H \bar U^2
    \right)
    +
    \frac{\partial~}{\partial Y}(\bar\rho H \bar U \bar V)
    \qquad\qquad
    \qquad\qquad
    \qquad\qquad
    \qquad\qquad
    \qquad\qquad\\
    \qquad+
        \frac{1}{\gamma}
        \left\{
            \left[
        1 + \left(\partial b/\partial Y\right)^2
        \right]\!
    \frac{\partial~}{\partial X}
    - \frac{\partial b}{\partial X}\frac{\partial b}{\partial
    Y}
    \frac{\partial~}{\partial Y}
        \right\}\!
    \left(
    \frac{\bar\rho g H^2}{2\gamma^2}
    \right)
    =
    -\frac{\bar\rho g H}{\gamma^2} \frac{\partial b}{\partial X}
    -T_{13},
    \label{eq:slope aligned mom 1}\end{split}\\
    \begin{split}%
    \frac{\partial~}{\partial t}(\bar\rho H \bar V)
    +
    \frac{\partial~}{\partial X}(\bar\rho H \bar U\bar V)
    +
    \frac{\partial~}{\partial Y}\!\left(
    \bar\rho H \bar V^2
    \right)
    \qquad\qquad
    \qquad\qquad
    \qquad\qquad
    \qquad\qquad
    \qquad\qquad\\
    \qquad+
        \frac{1}{\gamma}\left\{
    - \frac{\partial b}{\partial X}\frac{\partial b}{\partial
    Y}
    \frac{\partial~}{\partial X}
    +
        \left[
        1 + \left(\partial b/\partial X\right)^2
    \right]\!
    \frac{\partial~}{\partial Y}
        \right\}\!
    \left(
    \frac{\bar\rho g H^2}{2\gamma^2}
    \right)
    =
    -\frac{\bar\rho gH}{\gamma^2}\frac{\partial b}{\partial Y}
    -T_{23}.
     \label{eq:slope aligned mom 2}%
    \end{split}%
\end{gather}
\label{eq:slope aligned mom}%
\end{subequations}
The appearance of basal derivatives (via $\gamma$) in the hydrostatic pressure
and gravitational forcing terms of these equations stems from correctly
resolving these forces parallel to the slope.
As noted by Iverson \& Ouyang~\cite{Iverson2015}, these factors are often 
missing from Earth surface flow models, which can lead to significant errors
when steep slopes are present.
Furthermore, note that we made use of our assumption of negligible curvature, to
rearrange the hydrostatic pressure term and include an additional factor of
$1/\gamma$ within the gradient terms.  This is necessary to ensure that
Eqs.~\eqref{eq:slope aligned mom 1} and~\eqref{eq:slope aligned mom 2} preserve
\emph{lake-at-rest} steady states of the form
\begin{equation}
    \frac{H}{\gamma} + b \equiv \textrm{const}, \quad
    \bar{U}=\bar{V}
    =0,
    \quad \bar{\rho} \equiv \textrm{const},
    \label{eq:lake at rest analytic}
\end{equation}
which correspond to motionless flows with a horizontal free surface.  We assume
such states to be drag free and non-morphodynamic. (Purely granular flows are
not generally expected to achieve this state for example, since they may arrest
in non-flat, slumped configurations, due to static basal friction, see e.g.\
Ref.~\cite{Kerswell2005}.) After substituting, it may be easily verified that
Eq.~\eqref{eq:lake at rest analytic} is a solution to Eqs.~\eqref{eq:slope
aligned mom 1} and~\eqref{eq:slope aligned mom 2}.  Even so, it is nontrivial to
design a numerical scheme that preserves these steady states at finite
resolution. We show how to do so for our model equations in
Sec.~\ref{sec:numerics}.

Equations~(\ref{eq:shallow mass eqs}\emph{a,b}) and~(\ref{eq:slope aligned mom}\emph{a,b}), plus the bed evolution equation~\eqref{eq:bed evolution}, comprise
a complete set of shallow-layer governing equations for the flow.  Returning to
the coordinate transformation in Eq.~\eqref{eq:xyz}, we see that after
depth-averaging, the surface-fitted coordinates collapse to a Cartesian
description where $x\equiv X$ and $y \equiv Y$. Denoting the depth-averaged
velocity vector in the Cartesian basis $\{\vect{e}_x,\vect{e}_y,\vect{e}_z\}$ as
$\bar{\vect{u}}=(\bar u, \bar v, \bar w)^\tran$, it is straightforward to see
that $\bar u = \bar U$ and $\bar v = \bar V$. The remaining component is given
by
\begin{equation}
    \bar w = \vect{e}_z \cdot (\bar U \vect{e}_X + \bar V \vect{e}_Y) = 
    \bar u \frac{\partial b}{\partial x} + \bar v \frac{\partial b}{\partial
    y}.
    \label{eq:wbar}%
\end{equation}
Therefore, for the remainder of the paper, we use Cartesian coordinates, since
these are convenient for numerical simulations of Earth-surface flows, which are
typically run on digital elevation data parametrised in a gravity-aligned frame.
However, the curvilinear transformation above was nevertheless required in order
to depth integrate the flow along the surface normal and retain the correct
geometrical dependence in the model equations.  Furthermore, note that since we
need not project $H$ onto the vertical when transforming back to Cartesian, it
still carries the correct physical meaning of the flow depth measured along the
bed normal.

\subsection{Regularisation}
The characteristics $\lambda_1$ and $\lambda_2$ of our morphodynamic shallow
layer model, restricted to the $x$ and $y$ directions
respectively, are
\begin{subequations}
\begin{equation}
    \lambda_1 = \bar{u} \pm \gamma^{-3/2}\sqrt{gH(1+b_y^2)}, \bar{u}, \bar{u}, 0
    ~\quad\textrm{and}\quad~
    \lambda_2 = \bar{v} \pm \gamma^{-3/2}\sqrt{gH(1+b_x^2)}, \bar{v}, \bar{v}, 0,
    \tag{\theequation\emph{a,b}}%
\end{equation}
    \label{eq:characteristics}%
\end{subequations}
where $b_x \equiv \partial b / \partial x$ and $b_y \equiv \partial b/\partial
y$.  The eigenspace associated with these characteristics is degenerate when
either 
\begin{subequations}
\begin{equation}
    \gamma^{3/2} \bar{u} = \sqrt{g H (1 + b_y^2)}
    ~\quad\textrm{or}\quad~
    \gamma^{3/2} \bar{v} = \sqrt{g H (1 + b_x^2)},
    \tag{\theequation\emph{a,b}}%
\end{equation}
\end{subequations}
since it may be verified that two of the characteristics are zero and both
possess the same eigendirection.
This implies that the governing equations
are not everywhere hyperbolic and consequently ill posed as an initial value
problem~\cite{Joseph1990}. This agrees with the more detailed analysis of
Ref.~\cite{Langham2021}, which considers a system that is structurally very
similar. Therefore, it is essential to regularise the governing equations to
obtain well-posed models. Following Ref.~\cite{Langham2021}, we 
add the terms
\begin{equation}
    \frac{\partial~}{\partial x} \left( \nu \bar{\rho} H \frac{\partial
    \bar{u}}{\partial x} \right)+
    \frac{\partial~}{\partial y} \left( \nu \bar{\rho} H \frac{\partial
    \bar{u}}{\partial y} \right)
    ~\quad\textrm{and}\quad~
    \frac{\partial~}{\partial x} \left( \nu \bar{\rho} H \frac{\partial
    \bar{v}}{\partial x} \right)+
    \frac{\partial~}{\partial y} \left( \nu \bar{\rho} H \frac{\partial
    \bar{v}}{\partial y} \right)
    \label{eq:diffusion}%
\end{equation}
to the right-hand sides of the momentum equations~\eqref{eq:slope aligned mom 1}
and~\eqref{eq:slope aligned mom 2} respectively, which model the diffusion of
momentum via turbulent eddies.  The free parameter $\nu$ controls the magnitude
of these terms, which are typically expected to be small, relative to other
contributions to the flow momentum, since turbulent dissipation occurs at scales
smaller than the flow depth.
Consequently, it is likely to have only a minimal effect on global flow
dynamics. Nevertheless, its inclusion has the desired effect of damping out
unphysical resonances that otherwise occur at short wavelengths when
the characteristics coalesce.

An illustration of this regularisation is given in Fig.~\ref{fig:ill
posedness}(\emph{a}), using simulations of the governing equations restricted to
one spatial dimension. 
\begin{figure}
    \begin{centering}%
    \includegraphics[width=\textwidth]{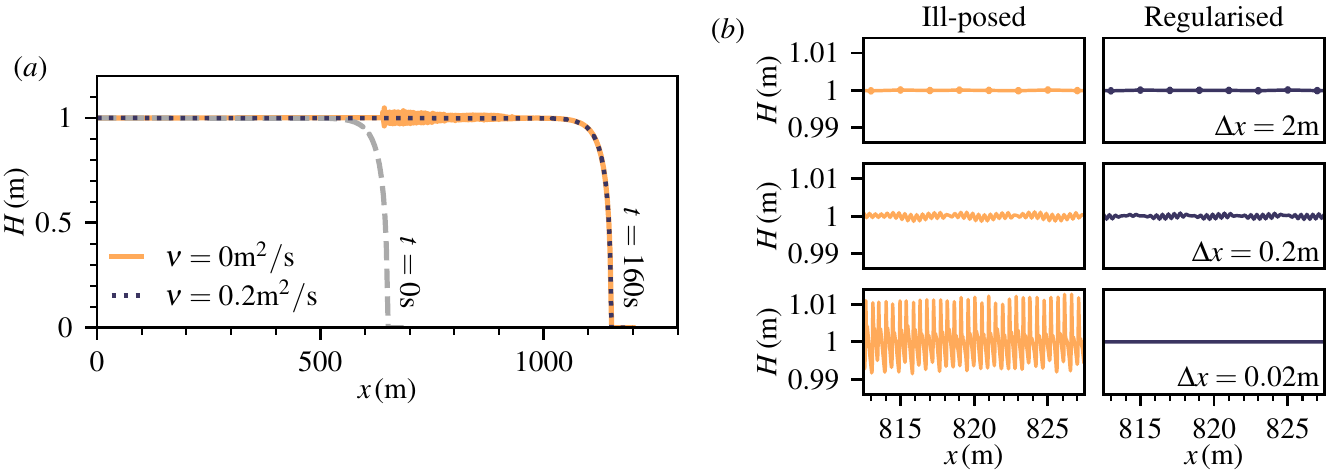}%
        \caption{%
        Demonstration of the model regularisation.
        (\emph{a})~The flow depth 
        as a function of distance, for simulations 
        with $\nu =
        0\textrm{m}^2/\textrm{s}$ (solid yellow) and $\nu =
        0.2\textrm{m}^2/\textrm{s}$ (dotted purple) at $t=160\textrm{s}$.  The
        grid resolution is $\Delta x = 0.02\textrm{m}$. The initial condition
        ($t=0\textrm{s}$), plotted in dashed grey, is a steady travelling wave
        in the governing system without morphodynamics ($M \equiv 0$) on a
        constant slope, whose gradient ($0.04$) was
        selected so that $\bar{u} = \sqrt{gH}/\gamma^{3/2}$ in the uniform tail.
        The solution was obtained using the numerical method presented
        in Sec.~\ref{sec:numerics} with the parameters and example closures
        given in Sec.~\ref{sec:results}, by numerically
        integrating flow out from a Dirichlet boundary that imposes
        $H=1\textrm{m}$, $\bar u = 3.128\textrm{m/s}$~(4~s.~f.) and
        $\bar{\psi}=2.295\times 10^{-2}$~(4~s.~f.) at $x=0\textrm{m}$. [The
        condition on $\bar u$ is the velocity of the corresponding steady
        uniform layer, which may be calculated exactly via Eq.~(\ref{eq:steady
        balances}\emph{a}). The condition on $\bar\psi$ is the corresponding
        solids fraction at which the bed evolution equation possesses a stable
        equilibrium and may be computed by finding the dilute root of $M(H,\bar
        u, \bar\psi) = 0$, given the closures in
        Eqs.~(\ref{eq:M})--(\ref{eq:EandD}\emph{b}).] (\emph{b})~Close-up views of the
        free surface for corresponding simulations with three different grid
        resolutions, as labelled on the right-hand panels.  The left-hand panels
        use $\nu = 0\textrm{m}^2/\textrm{s}$ and the right-hand panels use $\nu =
        0.2\textrm{m}^2/\textrm{s}$.  In the top row only, filled circles show
        the locations of numerical grid points.
        }
    \label{fig:ill posedness}%
    \end{centering}%
\end{figure}
This shows the evolution of a travelling wave on an initially constant slope,
prepared such that its uniform tail lies at the transcritical point where the
second and fifth characteristics in Eq.~(\ref{eq:characteristics}\emph{a})
intersect.  With no regularisation~($\nu = 0\textrm{m}^2/\textrm{s}$), the
resulting motion is unstable and leads to extremely rapid spatial oscillations,
emanating from the front region, which is not initially in morphodynamic
equilibrium and acts as a perturbation to the steady tail.  These oscillations
have approximately zero wave speed and appear as `noise' in the plotted profile.
Applying a sufficiently high eddy viscosity ($\nu = 0.2\textrm{m}^2/\textrm{s}$)
stabilises the flow. In principle, it is difficult to constrain $\nu$ to a
particular constant value, since the amount of turbulence dissipation is \emph{a
priori} unknown and depends on the spatiotemporal flow dynamics.  However, our
choice here is commensurate with the hypothesis that the scale of eddy
viscosity should depend on the product of the flow depth with the turbulent
shear velocity $u_\star = \sqrt{\tau_b / \bar\rho}$, where~$\tau_b$ denotes the
magnitude of basal drag (assuming the model closures and parameter choices of
Sec.~\ref{sec:results}).  Since it is not our intention to examine the effect of
$\nu$ in detail, it suffices to illustrate our results herein.

In Fig.~\ref{fig:ill posedness}(\emph{b}), we show the effect of changing the
length scale of the numerical discretisation~$\Delta x$ on a small patch of the
free surface. The left-hand panels show the unregularised solution at
$t=160\textrm{s}$, for the three resolutions labelled on the right-hand panels,
which show the corresponding regularised snapshot. The coarsest results use
$\Delta x = 2\textrm{m}$, which is broadly representative of resolutions used in
operational shallow flow simulations~\cite{Liu2013,Barnhart2021,Jenkins2023}.
Here, there is little difference between the regularised and unregularised data,
which is nearly completely level.  In this case, the length scale of the
instability is smaller than $\Delta x$, so it cannot be captured by the numerics
in either case.  Corresponding results are shown at $\Delta x = 0.2\textrm{m}$
and $0.02\textrm{m}$. Refinement introduces short unstable wavelengths into the
numerical solutions. In the ill-posed case, finer resolution can only lead to
increasingly rapid unstable growth, further polluting the free surface.
Conversely, the regularised results eventually converge to a level profile.
Though the converged profile does ultimately agree with the ill-posed result at
$2\textrm{m}$ resolution (due to the effects of numerical diffusion), we caution
that in the ill-posed case, such results are simply not robust enough to be
relied upon for flow prediction.

\section{Numerical method}
\label{sec:numerics}%
In this section, we present an approach for solving the flow
equations~(\ref{eq:shallow mass eqs}\emph{a,b}),~(\ref{eq:slope aligned
mom}\emph{a,b}), augmented by the regularising terms~\eqref{eq:diffusion},
together with Eq.~\eqref{eq:bed evolution} for the bed evolution.  Our primary
focus is on adapting existing finite volume shallow water schemes to incorporate
morphodynamics with a suspended sediment phase and featuring the geometric
corrections derived in~\S\ref{sec:derivation}, rather than on constructing a
bespoke method.

We firstly write the flow equations in vector form, decomposing them into their
hydraulic and morphodynamic components. The mass conservation equations
[Eqs.~(\ref{eq:shallow mass eqs}\emph{a,b})] are
linearly recombined to give equations for $H$ and $\bar\psi H$.
Then we define a vector of primary observables
$\vect{q}\equiv (H, \bar{\psi} H, \bar{\rho}H\bar{u},
\bar{\rho}H\bar{v})^\tran$, and a
two-dimensional gradient operator $\nabla \equiv (\partial / \partial x,
\partial / \partial y)^\tran$.
Letting $O_n$ and $I_n$ denote the
$n\times n$ zero and identity matrices respectively,
the governing system is then given by
\begin{subequations}
\begin{gather}
    \frac{\partial \vect{q}}{\partial t}
    = \mathcal{H}(\vect{q}) + \mathcal{M}(\vect{q}),\label{eq:dqdt partial}\\
    \frac{\partial b}{\partial t} = -\gamma M(\vect{q}),
    \label{eq:bed evolution numeric}%
\end{gather}
\end{subequations}
where
\begin{subequations}
\begin{gather}
    \mathcal{H}(\vect{q}) = 
    -\frac{\partial \vect{F}_1(\vect{q})}{\partial x}
    -\frac{\partial \vect{F}_2(\vect{q})}{\partial y}
    -
    \gamma B \nabla \zeta(\vect{q})
    + \vect{S}_1(\vect{q})
    + \vect{S}_2(\vect{q}),\label{eq:hydraulic op}\\
    \mathcal{M}(\vect{q}) = \left[\morpho(\vect{q}), \psi_b \morpho(\vect{q}),
    0, 0\right]^{\,\tran}\!\!,
\end{gather}
\end{subequations}
with $\zeta = \frac{1}{2}\bar{\rho}g H^2/\gamma^2$,
\begin{subequations}
\begin{gather}
    \vect{F}_1(\vect{q}) = 
    \left( H\bar{u}, \bar{\psi} H \bar{u}, \bar{\rho} H \bar{u}^2, \bar{\rho} H
    \bar{u}\bar{v}
    \right)^\tran\!\!, \quad
    \vect{F}_2(\vect{q}) = 
    \left( H\bar{v}, \bar{\psi} H \bar{v}, \bar{\rho} H \bar{u}\bar{v}, \bar{\rho} H
    \bar{v}^2
    \right)^\tran\!\!,\tag{\theequation\emph{a,b}}\\
    \label{eq:B}%
    B = \begin{pmatrix}
        O_2 \\
        I_2 - \gamma^{-2}\nabla b \otimes \nabla b
    \end{pmatrix},\quad
    \vect{S}_1(\vect{q}) =
    \left[
        0, 0, -\frac{\bar{\rho}g H}{\gamma^2}\frac{\partial b}{\partial x}
        -\tau_{13}(\vect{q}), -\frac{\bar{\rho}g H}{\gamma^2}\frac{\partial
        b}{\partial y} -\tau_{23}(\vect{q})
    \right]^\tran\!\!\!,\tag{\theequation\emph{c,d}}\\
    \vect{S}_2(\vect{q}) = \left[
    0, 0, 
	\nabla\cdot \left(\nu \bar\rho H \nabla \bar u\right),
	\nabla\cdot \left(\nu \bar\rho H \nabla \bar v\right)
    \right]^\tran\!\!.\tag{\theequation\emph{e}}
\label{eq:S2}%
\end{gather}
    \label{eq:FBS}%
\end{subequations}

The vector $\vect{q}$ is discretised spatially over a regular Cartesian mesh by averaging
its components over rectangular cells,
$c_{i,j} \equiv [x_{i-1/2},x_{i+1/2}] \times [y_{j-1/2},y_{j+1/2}]$, defining
\begin{equation}
    \vect{q}_{i,j}(t) = \frac{1}{\Delta x \Delta y} \int_{c_{i,j}}
    \vect{q}(x,y,t) \,\mathrm{d}x\,\mathrm{d}y,
    \quad
    \textrm{for}~i,j \in \mathbb{Z},
\end{equation}
where $\Delta x$ and $\Delta y$ are fixed grid spacings such that
$x_{i+1/2}-x_{i-1/2}=\Delta x$ and $y_{j+1/2}-y_{j-1/2}=\Delta y$.  We 
denote the centroid of $c_{i,j}$ by $(x_i, y_j)$.  The full numerical solution
vector $\widetilde{\vect{q}}$ (at each time $t$) is the collection of all the
$\vect{q}_{i,j}$ for each cell in the simulation domain.
The temporal evolution of the numerical problem obeys a system of ordinary
differential equations~(ODEs) encapsulated by
\begin{equation}
    \frac{\mathrm{d} \widetilde{\vect{q}}}{\mathrm{d} t}
    = \widetilde{\mathcal{H}}(\widetilde{\vect{q}}) + 
    \widetilde{\mathcal{M}}(\widetilde{\vect{q}}),
    \label{eq:dqdt}%
\end{equation}
where $\widetilde{\mathcal{H}}$ and $\widetilde{\mathcal{M}}$ are suitably
chosen discrete approximations to the partial differential operators
$\mathcal{H}$ and $\mathcal{M}$ respectively. In particular,
$\widetilde{\mathcal{H}}$ is responsible for numerically evaluating the spatial
derivatives present in the flux, gradient and diffusive terms of
Eq.~\eqref{eq:hydraulic op}.

Likewise, the bed equation may be spatially discretised by evaluating
Eq.~\eqref{eq:bed evolution numeric} over a lattice of points. It is coupled to
Eq.~\eqref{eq:dqdt} because the morphodynamics closure $\morpho$ depends on
$\vect{q}$. A convenient way to incorporate this coupling within an existing
shallow layer finite volume method is to employ operator splitting to integrate
Eq.~\eqref{eq:dqdt}.  This enables the morphodynamics to be time stepped
separately from the flow, leaving the original scheme unmodified. Moreover, we
find that this approach simplifies the job of ensuring desirable properties such
as mass conservation and steady state preservation, to be discussed shortly.
Operator splitting has also been employed recently by Chertock \emph{et
al.}\ for simulating the related (but inequivalent) physical problem of
shallow water flows coupled with bed load transport~\cite{Chertock2020}.

We propose that the governing equations are evolved in time using the
commonplace second-order splitting due to
Strang~\cite{Strang1968}, which updates the gridded solution data 
via the following decomposition
\begin{equation}
    \widetilde{\vect{q}}(t + \Delta t) = 
    \widetilde{\mathcal{H}}_{\Delta t / 2} \widetilde{\mathcal{M}}_{\Delta t} 
    \widetilde{\mathcal{H}}_{\Delta t / 2} \left[\widetilde{\vect{q}}(t)\right].
    \label{eq:op splitting}%
\end{equation}
Here, $\widetilde{\mathcal{H}}_{\Delta t}$ and $\widetilde{\mathcal{M}}_{\Delta t}$
represent time-discretised analogues of $\widetilde{\mathcal{H}}$ and
$\widetilde{\mathcal{M}}$, which approximate the evolution of 
the respective subproblems
\begin{subequations}
    \label{eq:subproblems}%
\begin{gather}
    \frac{\mathrm{d} \widetilde{\vect{q}}}{\mathrm{d} t}
    = \widetilde{\mathcal{H}}(\widetilde{\vect{q}}), \quad
    \frac{\mathrm{d} \widetilde{\vect{q}}}{\mathrm{d} t}
    =
    \widetilde{\mathcal{M}}(\widetilde{\vect{q}}),
    \tag{\theequation\emph{a,b}}%
\end{gather}
\end{subequations}
over an interval $\Delta t$ forward in time using appropriately chosen time
stepping algorithms.  This approach effectively decouples the hydraulic and
morphodynamic physics from a numerical point of view. The bed equation need only
be solved in concert with the morphodynamic part, while the implementation of
the hydraulic operator may assume that the bed surface is static.
Therefore, we split the remainder of this section into separate discussions of
the hydraulic and morphodynamic numerics.

\subsection{Hydraulic step}
The hydraulic component of our governing equations involves mostly standard, or
slightly modified versions of terms found in classical shallow water
formulations.  Consequently, the evaluation of $\widetilde{\mathcal{H}}_{\Delta
t}$ may be achieved by extending existing schemes that are well documented
elsewhere~\cite{LeVeque2002,Kurganov2002,Kurganov2018}.  We base our particular
implementation on the central-upwind finite volume solver of Chertock \emph{et
al}.~\cite{Chertock2015a}, which possesses desirable qualities, including the
exact preservation of steady states (including lake-at-rest) and guaranteed
positivity of the flow depth. Since the full details of this method are
not our focus, we concentrate only on describing the specific adaptations that
are needed to solve our model. We have endeavoured to keep our framing as
general as possible, so that these adaptations can be used to augment other
shallow water schemes.

At any point in the simulation, each of the flux terms within
$\widetilde{\mathcal{H}}$ may be evaluated on $\widetilde{\vect{q}}$ at
$(x_i,y_j)$, using numerical derivatives of the form
\begin{subequations}
\begin{gather}
    \frac{\partial f}{\partial x}(x_i, y_j) \approx
    \frac{f(\vect{q}_{i+1/2,j}) 
    - 
    f(\vect{q}_{i-1/2,j})}
    {\Delta x}, \quad
    \frac{\partial f}{\partial y}(x_i, y_j) \approx
    \frac{f(\vect{q}_{i,j+1/2}) 
    - 
    f(\vect{q}_{i,j-1/2})}
    {\Delta y},
    \tag{\theequation\emph{a,b}}
    \label{eq:dfdx dfdy}%
\end{gather}
\end{subequations}
where $f$ is a field in $\vect{F}_1$ or $\vect{F}_2$ respectively.
Here, we abuse notation slightly: $f(\vect{q}_{i\pm1/2,j})$ and
$f(\vect{q}_{i,j\pm 1/2})$ should be understood as numerical fluxes, i.e.\
approximations to the average value of $f$ evaluated across the cell interfaces
$\{x_{i\pm1/2}\}\times [y_{j-1/2},y_{j+1/2}]$ and
$[x_{i-1/2},x_{i+1/2}]\times\{y_{j\pm 1/2}\}$ respectively.
Various strategies exist for the computation of numerical fluxes, the details of
which are fairly complicated and not especially important here.  Unfamiliar
readers may refer to e.g.\ Ref.~\cite{LeVeque2002} for an overview. 
The diffusive terms in $\vect{S}_2$ may be
computed within the framework of numerical fluxes, or via finite difference
stencils, as in Ref.~\cite{Chertock2008} for example. In our implementation, to compute
the first term of diffusion in the $x$-direction, we employ a variation on an approach documented
by Kurganov and Tadmor~\cite{Kurganov2000}, setting $f = \nu \bar \rho H
\partial \bar u/\partial x$ and computing the corresponding numerical flux as
the average of two approximate values either side of the relevant interface via
the general formula:
\begin{equation}
f(\vect{q}_{i+1/2,j}) =
\frac{1}{2} \left[
    f\left(\vect{q}_{i+1/2,j}^-, (\vect{q}_x)_{i,j}\right) +
    f\left(\vect{q}_{i+1/2,j}^+, (\vect{q}_x)_{i+1,j}\right)
\right].
\label{eq:Px}%
\end{equation}
Here, $\vect{q}_{i+1/2,j}^- = \vect{q}_{i,j} + \Delta x (\vect{q}_x)_{i,j}/2$,
$\vect{q}_{i+1/2,j}^+ = \vect{q}_{i+1,j} - \Delta x (\vect{q}_x)_{i+1,j}/2$, are
reconstructions of~$\widetilde{\vect{q}}$ at either side of the corresponding
cell interface, using $(\vect{q}_x)_{i,j}$ and $(\vect{q}_x)_{i+1,j}$, which
denote numerical approximations to $\partial \vect{q}/\partial x$ at $(x_i,y_j)$
and $(x_{i+1},y_j)$ respectively. These latter terms are routinely computed as
part of central-upwinding schemes (separate from, but in conjunction with the
flux derivatives themselves) by using flux-limiting formulae. Computation of
$\bar \rho_{i,j}$, which is required, but not explicitly available in
$\vect{q}_{i,j}$ is discussed shortly.
Analogous formulae to Eq.~\eqref{eq:Px} may be applied to compute the remaining
components of the Laplacian.

In the case of evaluating the hydrostatic pressure term, $\gamma B\nabla\zeta$,
particular attention must be paid to the problem of preserving
lake-at-rest solutions, defined above in Eq.~\eqref{eq:lake at rest analytic}
to be stationary states with a horizontal free surface.  Schemes that do not
compute these states exactly generate spurious momentum from static initial
conditions. This is undesirable for simulations of Earth surface flows that can
naturally arrest. 
It is easily verified from the governing equations that numerically preserving
these states requires that in the resting configuration [Eq.~\eqref{eq:lake at
rest analytic}], the discretised hydrostatic pressure gradient and
gravitational forcing exactly balance.
Recall that we do not include rheologies that exert nonzero basal stresses when
resting in these considerations, since they may adopt a non-horizontal free
surface when at rest.

To achieve this, we make use of a second-order discretisation due to Kurganov
and Levy~\cite{Kurganov2002} that computes these terms using values for the flow
surface $\eta \equiv H/\gamma + b$ at cell centres and the bed height $b$ at
cell vertices.  Consequently, we store and update $b$ at the cell vertices
$(x_{i\pm1/2},y_{j_\pm1/2})$ and use linear interpolation to obtain the
necessary topographic data at cell interfaces 
\begin{subequations}
\begin{equation}
    b_{i\pm1/2,j} = \frac{1}{2}\left(b_{i\pm1/2,j-1/2} +
    b_{i\pm1/2,j+1/2}\right)\!,~~
    b_{i,j\pm1/2} = \frac{1}{2}\left(b_{i-1/2,j\pm2} + b_{i+1/2,j\pm2}\right)
    \label{eq:b ifaces}%
    \tag{\theequation\emph{a,b}}%
\end{equation}
\end{subequations}
and centres
\begin{subequations}
\begin{equation}
    b_{i,j} = \frac{1}{2}\left(
    b_{i-1/2,j} +
    b_{i+1/2,j}
    \right)\!, 
    ~
    (b_x)_{i,j} = \frac{b_{i+1/2,j} - b_{i-1/2,j}}{\Delta x},
    ~
    (b_y)_{i,j} = \frac{b_{i,j+1/2} - b_{i,j-1/2}}{\Delta y}.
    \label{eq:b centre}%
    \tag{\theequation\emph{a--c}}%
\end{equation}
    \label{eq:b centre specific}%
\end{subequations}
Then, for the hydrostatic pressure gradient, we write $\zeta =
\frac{1}{2}\bar{\rho}g(\eta-b)^2$ and compute
\begin{equation}
    \gamma(x_i,y_j) B(x_i,y_j) \left[
        \frac{\partial \zeta}{\partial x}(x_i,y_j),
        \frac{\partial \zeta}{\partial y}(x_i,y_j)
        \right]^\tran\!\!,
    \label{eq:hydrostatic pressure discrete}%
\end{equation}
evaluating the geometric terms $\gamma$, $B$ at cell centres, using 
Eqs.~\eqref{eq:b ifaces} and~\eqref{eq:b centre} with their respective formulae
in Eqs.~\eqref{eq:n} and~(\ref{eq:FBS}\emph{c}),
and evaluating the $\zeta$ derivatives via suitable numerical fluxes as
per the discussion around Eqs.~\eqref{eq:dfdx dfdy}.
The gravitational forcing term is similarly evaluated at each cell centre. Its
discretisation, by way of Eqs.~\eqref{eq:b centre}, is
\begin{equation}
    -\frac{\bar{\rho}_{i,j} g (\eta_{i,j} - b_{i,j})}{\gamma(x_i,y_j)} \left[ (b_x)_{i,j}, (b_y)_{i,j}
    \right]^\tran\!.
    \label{eq:grav forcing discrete}%
\end{equation}
In both Eqs.~\eqref{eq:hydrostatic pressure discrete} and~\eqref{eq:grav forcing
discrete}, $\bar{\rho}_{i,j}$ must be determined from $\vect{q}_{i,j}$. It may
be computed via Eq.~\eqref{eq:rho}, after obtaining $\bar{\psi}_{i,j}$ from
$(\bar\psi H)_{i,j}$, by using an appropriate formula that guards against
numerical singularities when flow depths are small~\cite{Chertock2015a}, such as
\begin{equation}
    \bar{\psi}_{i,j} = \frac{2H_{i,j}(\bar\psi H)_{i,j}}{H_{i,j}^2 +
    \max\{H_{i,j}^2,H_\epsilon^2\}},
\end{equation}
where $H_\epsilon$ is a small constant. In our implementation, we use
$H_\epsilon = 10^{-6}\textrm{m}$.  Analogous formulae may deployed (when required) to obtain
the flow velocity from $(\bar\rho H\bar u)_{i,j}$ and $(\bar\rho H\bar
v)_{i,j}$. 

It may now be verified that the numerical terms in Eqs.~\eqref{eq:hydrostatic
pressure discrete} and~\eqref{eq:grav forcing discrete} exactly balance for
lake-at-rest flows. The presence of the matrix $B$ [defined in
Eq.~(\ref{eq:FBS}\emph{c})] makes confirming this fact
slightly more complicated than in prior instances of schemes for the classical
shallow water equations~\cite{Kurganov2002,Kurganov2007,Chertock2015a}. We show it
explicitly in Appendix~\ref{appendix:lar}.

The remaining terms in $\widetilde{\mathcal{H}}$ yet to be discussed are the
cell-averaged components of the bottom drag closures, $\tau_{13}(\vect{q})$ and
$\tau_{23}(\vect{q})$.  Regardless of the choice of model parametrisation, these
should not be evaluated naively on $\vect{q}_{i,j}$, which only contains the
horizontal components of momentum.  For simplicity, assume that the drag force
density vector depends isotropically on the flow velocity, i.e.\ it has
magnitude $\tau_b \equiv \tau_b(|\bar{\vect{u}}|; H, \bar{\psi})$. Then, at each
cell $c_{i,j}$, we must evaluate
\begin{equation}
    (\tau_{13}, \tau_{23})_{i,j} =
    \tau_b(|\bar{\vect{u}}_{i,j}|)\frac{(\bar{u}_{i,j},\bar{v}_{i,j})}{|\bar{\vect{u}}_{i,j}|},
    ~~
    \textrm{where}
    ~~
    |\bar{\vect{u}}_{i,j}| = 
    \sqrt{\bar{u}_{i,j}^2 + \bar{v}_{i,j}^2 + 
    \left(\bar{u}_{i,j}(b_x)_{i,j} + \bar{v}_{i,j}(b_y)_{i,j}\right)^{\!2}}.
    \label{eq:drag evaluation}%
\end{equation}
This formula ensures that the contribution of the vertical velocity
[Eq.~\eqref{eq:wbar}] is consistently transmitted to the momentum balance by
accounting for the underlying basal geometry. 

Furnished with the means to compute the various components of
$\widetilde{\mathcal{H}}(\widetilde{\vect{q}})$, the hydraulic subproblem
[Eq.~(\ref{eq:subproblems}\emph{a})] may be time stepped from a given initial
condition, using any number of suitable methods.  In our implementation, we
follow Chertock \emph{et al.}~\cite{Chertock2015a} and use a second-order
semi-implicit Runge-Kutta scheme, detailed in Ref.~\cite{Chertock2015b}. To
construct a scheme that is mass conserving, we must be observant of the
geometry of the system. The total volume of mixed material and of solids may be
obtained by integrating over the flow and recalling that the Jacobian determinant
of the transformation to slope-fitted coordinates is $\gamma$ (up to order
$\epsilon^2$). These are given by
\begin{equation}
\int_{\mathbb{R}^2} H \gamma \,\mathrm{d}x\,\mathrm{d}y
\quad
\textrm{and}
\quad
\int_{\mathbb{R}^2} \bar\psi H \gamma \,\mathrm{d}x\,\mathrm{d}y,
\end{equation}
respectively. Expressions for the total mass of fluids and solids are linearly
dependent on these integrals. Therefore, to conserve these quantities during
the hydraulic step, it is sufficient to conserve the fluxes of $H\gamma$ and
$\bar\psi H\gamma$ across numerical cells. This is achieved by updating
$H_{i,j}$ over the time interval $\Delta t$ as so
\begin{gather}
    \begin{split}
H_{i,j}(t + \Delta t) - H_{i,j}(t) = 
\quad\quad
\quad\quad
\quad\quad
\quad\quad
\quad\quad
\quad\quad
\quad\quad
\quad\quad
\quad\quad
\hspace{3cm}
\\
\quad\quad
-\frac{\Delta t}{\gamma_{i,j}}\left\{
\left[
\frac{(H\bar u \gamma)_{i+1/2,j}}{\Delta x} -
\frac{(H\bar u \gamma)_{i-1/2,j}}{\Delta x}
\right]
+
\left[
\frac{(H\bar v \gamma)_{i,j+1/2}}{\Delta y} -
\frac{(H\bar v \gamma)_{i,j-1/2}}{\Delta y}
\right]\right\},
    \end{split}
    \label{eq:H update}%
\end{gather}
where $\gamma_{i,j}=\gamma(x_i,y_j)$. The numerical fluxes on the right-hand
side of Eq.~\eqref{eq:H update} should be understood as being evaluated via an
appropriate method, as indicated in the discussion around Eq.~\eqref{eq:dfdx
dfdy}. Regardless of how they are approximated, the sum of these fluxes over the
whole simulation domain is zero, so $H\gamma$ is conserved between time steps.
An analogous formula may be used to update $(\bar\psi H)_{i,j}$.

In order to evolve $\eta = H/\gamma + b$, we take advantage of the fact that within the hydraulic
operator, $\partial \eta/\partial t = \gamma^{-1}\partial H/\partial t$. Therefore,
we may integrate each $\eta_{i,j}$ using the update 
to $H_{i,j}$, i.e.\ 
\begin{equation}
    \eta_{i,j}(t + \Delta t) = \eta_{i,j}(t)
    +
    \gamma_{i,j}^{-1} \left[
        H_{i,j}(t + \Delta t) -
        H_{i,j}(t)
        \right].
    \label{eq:w hydro update}%
\end{equation}
It is not strictly necessary to explicitly store $H_{i,j}$ in order to apply
this formula.  The steps in Eqs.~\eqref{eq:H update} and~\eqref{eq:w
hydro update} may be combined to time step the hydraulic subproblem
conservatively.
Schemes which make the small angle approximation
$\gamma = 1$ may fail to be truly mass conservative if they solve equations
which conserve the flow depth measured along the vertical.

\subsection{Morphodynamic step}
In the morphodynamic step, we first time step $b$.
Since the topographic data is discretised at cell vertices, we interpolate
$\widetilde{\vect{q}}$, computing
\begin{equation}
    \vect{q}_{i+1/2,j+1/2} = \frac{1}{4}\left(
    \vect{q}_{i,j} +
    \vect{q}_{i+1,j} +
    \vect{q}_{i,j+1} +
    \vect{q}_{i+1,j+1}
    \right)
\end{equation}
for each vertex point $(x_{i+1/2},y_{j+1/2})$. The bed may then be evolved by
time stepping the ODEs
\begin{equation}
    \frac{\mathrm{d}b_{i+1/2,j+1/2}}{\mathrm{d}t}
    =
    -\gamma_{i+1/2,j+1/2} \morpho\!\left(\vect{q}_{i+1/2,j+1/2}\right)\!.
\end{equation}
We compute the corresponding morphodynamic updates to the flow variables by
determining them directly from the changes in bed height over the time interval
$\Delta t$ (which, in the case of an explicit multi-step method should be
understood as the increment of a particular substep).

Conservation of mass requires that the total amount of material eroded from (or
deposited to) the bed is added to (or subtracted from) the flow.
Across cell $c_{i,j}$, 
this has volume
\begin{equation}
    \frac{\Delta x \Delta y}{4}
    \sum_{\hat{v}\in \hat{V}(c_{i,j})}\Delta b_{\hat{v}} =
    \Delta x \Delta y\Delta b_{i,j},
\end{equation}
where $\hat{V}(c_{i,j})$ denotes the set of $c_{i,j}$'s four vertices and
$\Delta b_{\hat v}$, $\Delta b_{i,j}$ are the changes in $b$ over the time
interval $\Delta t$, at the vertex $\hat{v}$ and cell $c_{i,j}$ respectively.
To ensure that the update is conservative, we can insist that this is equal and
opposite to the corresponding change to flow volume $H_{i,j}\gamma_{i,j}$, over
$c_{i,j}$.
In other words, we require
\begin{equation}
    \Delta b_{i,j} = 
-\left[(H\gamma)_{i,j}(t+\Delta t) 
- (H\gamma)_{i,j}(t)\right]\!.
\end{equation}
After rearranging for $(H\gamma)_{i,j}(t + \Delta t)$,
this implies that $\eta$ may be time stepped according to
\begin{equation}
    \eta_{i,j}(t+\Delta t) =
    \frac{1}{\gamma_{i,j}(t + \Delta t)^2}
    \left[
        (H\gamma)_{i,j}(t) - \Delta b_{i,j}
    \right]
    + b_{i,j}(t + \Delta t).
    \label{eq:w update}%
\end{equation}
Similarly, since we assume that the bed consists of saturated solids occupying
volume fraction $\psi_b$, the solids fraction of each cell may be updated by
\begin{equation}
    (\bar\psi H)_{i,j}(t + \Delta t) = 
    \frac{1}{\gamma_{i,j}(t + \Delta t)}
    \left[
    (\bar\psi H \gamma)_{i,j}(t)
    - \psi_b \Delta b_{i,j}
    \right]\!.
    \label{eq:psi update}%
\end{equation}

The steps above are sufficient to time step the morphodynamic subproblem in
Eq.~(\ref{eq:subproblems}\emph{b}), which completes the numerical scheme in
principle. However, there are various issues to be mindful of when implementing
the scheme in an operational code that are particular to morphodynamic flows.

\subsection{Practical considerations}
In operational debris flow simulations, where flows travel over great distances
(tens of kilometres) it is often practically necessary to limit the grid
resolution to achieve feasible computation times.  Therefore, we recommend
applying some consistency checks during the numerics.

The time step taken for the hydraulic subproblem should be adaptively limited by
the standard Courant-Friedrichs-Lewy (CFL) condition, which prevents information
from travelling farther than a fixed fraction of the grid
spacing~\cite{LeVeque2002,Chertock2015a}.  The system characteristics given in
Eq.~\eqref{eq:characteristics} (which differ from those of classical shallow
water equations) are used for this calculation.  
On very fine grids, the diffusive time scale 
$\min\{\Delta x^2 / \nu, \Delta y^2 / \nu\}$ may be comparable to the
hydraulic time step given by the CFL condition. It is necessary for stability to
limit the time step to the smaller of the two.
We suggest further verifying
\emph{a posteriori} that between morphodynamic steps
\begin{equation}
    |H_{i,j}(t + \Delta t) - H_{i,j}(t)| < \beta H_{i,j}(t),
\end{equation}
for all $c_{i,j}$ and some fixed $\beta < 1$, so that the relative change in
flow depth is not too great.  If this condition is violated, $\Delta t$ may be
refined and the time step recomputed.  Note that since the time step is shared
between the two subproblems in the splitting scheme [Eq.~\eqref{eq:op
splitting}], the prior hydraulic step must also be recomputed in this case.  We
have used $\beta = 0.1$ to compute the results herein.

Even if a very small time step is taken, it is possible when calculating $\Delta
b_{i,j}$, for a cell to deposit a larger volume of solids that it possesses
according to the value of $\bar{\psi}_{i,j}$.  The volume of flow that is
available to be deposited within a cell occupies a fraction
$\bar{\psi}_{i,j}/\psi_b$ of the volume.
Therefore `excess' deposition occurs at $c_{i,j}$ if
\begin{equation}
    \Delta b_{i,j} > \frac{(\bar\psi H \gamma)_{i,j}(t)}{\psi_b},
    \label{eq:excess dep}%
\end{equation}
which we can see from Eq.~\eqref{eq:psi update} would leave $(\bar\psi H)_{i,j}$
negative. As well as being unphysical, such a step can be disastrous for the
solver if it ultimately leads to the computation of negative flow depths.
Nevertheless, it is not always practical to refine $\Delta t$ and recompute the
morphodynamic step until no cell satisfies inequality~\eqref{eq:excess dep}.
Therefore, we opt to apply a correction to these cells.  Making such an
adjustment is complicated by the fact that $\Delta b_{i,j}$ is contingent on all
four of the bed updates at $c_{i,j}$'s vertices, in accordance with
Eqs.~\eqref{eq:b ifaces} and~(\ref{eq:b centre specific}\emph{a}). Each vertex
$\hat{v} \in \hat{V}(c_{i,j})$ may be `depositional' or `erosive' over $\Delta
t$, depending on the sign of $\Delta b_{\hat{v}}$. In order for
inequality~\eqref{eq:excess dep} to be satisfied, at least one of them must have
$\Delta b_{\hat{v}} > 0$.  We denote by ${\hat{V}}_D(c_{i,j}; t, \Delta t)$, the
set of vertices where the bed update at time $t$ is positive and correct only
these vertices.
Specifically, for each $\hat{v} \in \hat{V}_D(c_{i,j}; t, \Delta t)$ we adjust
the the bed height update from $\Delta b_{\hat{v}}$ to $\Delta b_{\hat{v}} (1 -
\delta_{\hat{v}})$, where (omitting time dependence)
\begin{equation}
    \delta_{\hat{v}} = \frac{4 \left[ \Delta b_{i,j} - 
    (\bar\psi H \gamma)_{i,j}/\psi_b\right]}{\sum_{\hat{v}\in
    \hat{V}_D(c_{i,j})} \Delta b_{\hat{v}}}.
    \label{eq:deltav}%
\end{equation}
Note that $\delta_{\hat{v}} > 0$, due to inequality~\eqref{eq:excess dep}. Moreover, by
consulting Eqs.~\eqref{eq:b ifaces} and Eq.~(\ref{eq:b centre specific}\emph{a}), we see that
\begin{equation}
    4 \Delta b_{i,j} = \sum_{\hat{v} \in \hat{V}(c_{i,j})} \Delta b_{\hat{v}}
    \leq \sum_{\hat{v} \in \hat{V}_D(c_{i,j})} \Delta b_{\hat{v}}.
\end{equation}
and conclude $\delta_{\hat{v}} \leq 1$. Since $\delta_{\hat{v}} \in (0, 1]$, the corrected bed
update at each vertex is always smaller (less depositional), but preserves its
original sign (it never becomes erosional).  After updating the vertices, the
flow variables $\eta$ and $\bar\psi H$ must be recomputed, via Eqs.~\eqref{eq:w
update} and~\eqref{eq:psi update}, for all cells $c_{k,l}$ such that $\hat{V}(c_{k,l})
\cap \hat{V}_D(c_{i,j})\neq \emptyset$.  Finally, we note that there may be
neighbouring cells that satisfy inequality~\eqref{eq:excess dep}. In this case,
a vertex may receive up to four corrections. Therefore, as successive updates
are applied, Eq.~\eqref{eq:deltav} must be computed from the numerical solution
that incorporates any updates already proposed for neighbouring cells. This
ensures that vertices which receive multiple corrections are not
`over-corrected'---i.e.\ the cumulative effect of multiple updates may still be
encapsulated as $\Delta b_{\hat{v}} \mapsto (1-\delta)\Delta b_{\hat{v}}$ for some $\delta \in
(0,1]$.

We suggest only applying this correction procedure to cells whose flow depths
are very low, so that any adjustments made are small compared with the total
flow volume. For our results in the next section, we make the correction on any
cell satisfying both inequality~\eqref{eq:excess dep} and
$H_{i,j} < H_\epsilon = 10^{-6}\textrm{m}$. If a deeper cell is found to
satisfy inequality~\eqref{eq:excess dep}, then we refine the time step and
recompute it. 

\section{Example solutions}
\label{sec:results}%
In order to demonstrate the scheme in operation, we present some results on two
simple initial topographies: a uniform slope and a surface that smoothly
transitions between two constant grades.
While
our governing equations and numerical methods are applicable for a wide range of
closures, spanning both granular and fluid-like flows, we employ simple
expressions for our illustrative results.
For $\tau_b$, we use a turbulent fluid (or Ch\'ezy) drag closure:
\begin{equation}
    \tau_b(|\bar{\vect{u}}|) = \bar{\rho} C_d |\bar{\vect{u}}|^2\!,
    \label{eq:chezy}%
\end{equation}
where $C_d$ is a positive coefficient. This is physically appropriate for flows
with low solids content and analytically convenient, due to its relative
simplicity.

The morphodynamic transfer term may be
thought of as a competition between rates of erosion $E \geq 0$ and deposition
$D \geq 0$, exchanging saturated granular material at the basal solids fraction
$\psi_b$. 
The magnitudes of $E$ and $D$ depend on the local flow conditions. We
write this as
\begin{equation}
    M(H,\bar{\vect{u}},\bar\psi) = \chi(H)\left[\frac{E(\bar{\vect{u}}) -
    D(\bar\psi)}{\psi_b}\right].
    \label{eq:M}%
\end{equation}
The function $\chi$ is included in order to damp morphodynamics out when
the flow is very shallow. We find this to be a practical necessity that prevents
the unphysical bulking of rapid, thin flows.
Therefore, we set
\begin{equation}
    \chi(H) = \frac{1}{2}\left\{
        1 + \tanh\left[
            a\log\left(\frac{H}{H_c}\right)
        \right]
        \right\},
    \label{eq:chi}%
\end{equation}
where $H_c$ is a characteristic length scale at which morphodynamic processes
decay with decreasing flow depth, and $a$ is a free parameter that dictates how
sharp the decay is. Note that $\lim_{H\to 0}\chi(H) = 0$ and $\chi(H) \approx
1$ when $H \gg H_c$. In all simulations, we set $H_c = d$ and $a = 10$. 

Two simple closures that capture the essential physics of erosion and
deposition are given by
    \begin{subequations}
\begin{equation}
    E(|\bar{\vect{u}}|)
    = \frac{\varepsilon\tau_b(|\bar{\vect{u}}|)}{\bar\rho
    u_p 
    }\quad\textrm{and}\quad
    D(\bar\psi) = w_s\bar\psi\left(1-\frac{\bar\psi}{\psi_b}\right)\!.
    \tag{\theequation\emph{a,b}}%
\end{equation}
    \label{eq:EandD}%
    \end{subequations}
Here, erosion is proportional to the basal friction with non-negative
dimensionless coefficient~$\varepsilon$, which controls the erodibility of the
substrate.  This is a simplified analogue of empirical formulae that predict
erosion rates according to power laws of dimensionless basal (Shields) stress,
beyond an incipient threshold value (taken to be zero
here)~\cite{Lajeunesse2010}.
The
normalising factor $u_p = (g'_\perp d)^{1/2}$ is a characteristic particle velocity,
where $g'_\perp = g(\rho_s/\rho_f - 1)/\gamma$ is the reduced gravity resolved
normal to the slope and $d$ is the diameter of solid particles.
The quadratic deposition function allows for the effects of hindered settling at
high solids concentrations. It phenomenologically captures the empirical formulae
of Richardson \& Zaki~\cite{Richardson1954} and others~\cite{Spearman2017}.  The
free parameter $w_s$ sets the characteristic settling velocity of particles.  We
fix illustrative parameter values throughout this section, chosen to be
representative of natural materials. These are given in
table~\ref{tab:params}. 
\begin{table}
    \caption{Parameter values used in the example simulations.} 
    \label{tab:params}
\begin{tabular}{cccccccc}
    \hline
    $C_d$ & $\psi_b$ & $\varepsilon$ & $d$\,(m) & $g$\,(m/s$^2$) & $\rho_f$\,(kg/m$^3$) & $\rho_s$\,(kg/m$^3$) & $w_s$\,(m/s)  \\
    \hline
    $0.04$ & $0.65$ & $2.5\times 10^{-3}$ & $5\times 10^{-3}$ & $9.81$ & $1000$ & $2000$ & $0.2$ \\\hline
\end{tabular}
\vspace*{-4pt}
\end{table}

In our simulations we use an equispaced numerical grid with $\Delta x = \Delta y$.
Each result below was computed at multiple resolutions in order to verify
that solution vectors on successively fine grids approach a limiting
state as $\Delta x \to 0$. 
However, since our primary intention is to demonstrate the numerical scheme in
operation, we are ultimately only concerned that presented solutions are
qualitatively converged, in the sense that any essential flow features that we
describe ought to be preserved under grid refinement.  For any flow variable
$f$,
we denote its corresponding numerical solution on the grid of width $\Delta x$,
by $f^{\Delta x}(t)$ and restrict our consideration of this vector to its
components $f_{i,j}^{\Delta x}(t)$ within the inundated region, whose grid
points are indexed by the set $I(t) = \{(i,j) : H_{i,j}^{\Delta x}(t) >
H_\epsilon \}$.  The discrepancy between two discrete solution fields at
different resolutions may be measured by projecting the finer solution onto the
coarser grid and computing the normalised residual of their difference (omitting
time dependence):
\begin{equation}
    R(\Delta x_1, \Delta x_2; f) = 
    \frac{\big|f^{\Delta x_1} - f^{\Delta x_2}\big|_{L^1}}{\big|f^{\Delta
    x_2}\big|_{L^1}},
    \quad
    \textrm{where}
    \quad
    \big| f^{\Delta x} \big|_{L^1} = \frac{1}{|I|} \sum_{(i,j)\in I}
    \big|f^{\Delta x}_{i,j}\big|.
\end{equation}
We have verified that $R(\Delta x, \Delta x / 2; f) < R(2\Delta x, \Delta x; f)$
and ensured that all the solutions presented satisfy $R(\Delta x, \Delta x / 2;
f) < R_\epsilon = 0.05$ for all $f \in \{H, H\bar \psi, \bar\rho H \bar u,
\bar\rho H \bar v, b-b_0\}$, where $b_0$ denotes the initial bed height. In
other words, halving the grid spacing changes none of the solution variables by
more than $5\%$ on average. For most of our results the convergence of each
field is closer to $1\%$. A single exception to this is the $\bar\rho H \bar v$
field of the Fig.~\ref{fig:const slope severe 1} result, which was far tougher
to resolve due to the presence of a roll wave instability, requiring us to relax
the convergence criterion to $R_\epsilon = 0.08$ in this case.  The grid
discretisation for each result is provided in the corresponding figure caption.

\subsection{Uniform slopes.}
\label{sec:constant slopes}%
On the following initial slope inclined at angle $\vartheta$,
\begin{equation}
    b(x,y,0) = -\tan(\vartheta)x,
\end{equation}
we simulate flows spreading from a constant dilute source flux $Q =
50\textrm{m}^3/\textrm{s}$ at $(x,y)=(0,0)$.  The source is computed by
adding $Q \Delta t /(\gamma^2_{i,j} N_s\Delta x \Delta y)$ to the $\eta$
time-stepping update in Eq.~\eqref{eq:w hydro update}, for each cell whose
centre lies within a radius of $10\textrm{m}$ from the origin, where $N_s$
denotes the number of such cells. Provided that the flow travels over a far
greater distance than the source radius, this procedure approximates a point
source and it is referred to as such below.

\subsubsection{Gentle slopes.}
\begin{figure}
    \begin{centering}%
    \includegraphics[width=\textwidth]{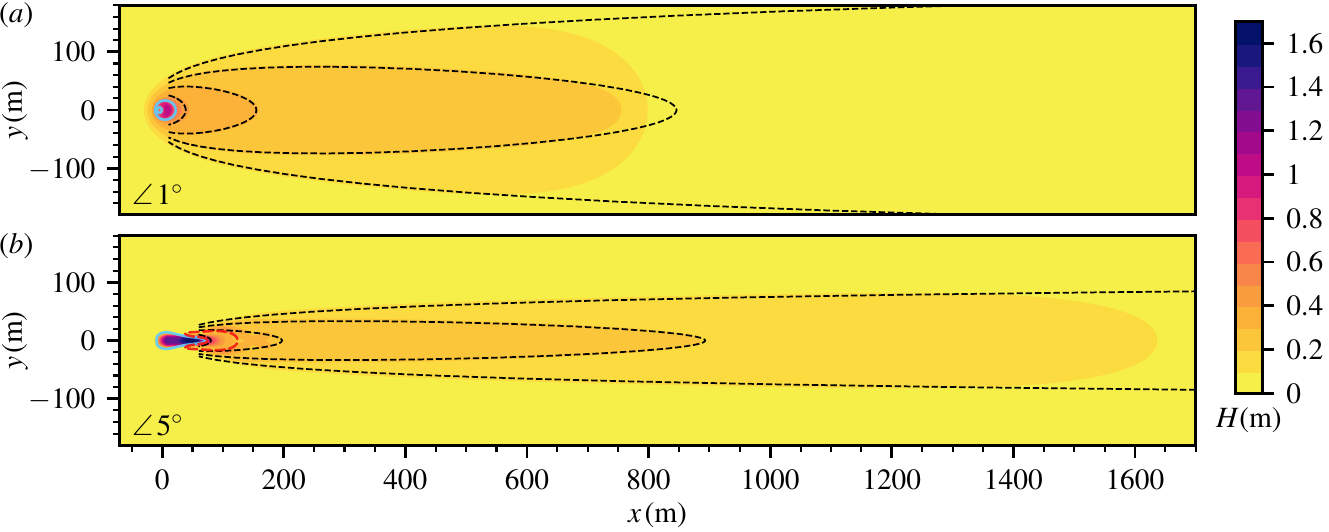}%
        \caption{
        Flows propagating on constant initial slopes of (\emph{a})~$1^\circ$ and
        (\emph{b})~$5^\circ$ inclination, fed by a $50\mathrm{m}^3/\mathrm{s}$
        point source flux at the origin.  Filled contours show the flow depth
        $H$ at $t = 800\mathrm{s}$ at increments of $0.1\mathrm{m}$.  Solid blue
        lines bound regions where the cumulative effect of morphodynamics has
        eroded the bed by more than $0.1\textrm{m}$. Likewise, the dashed red
        line in~(b) encloses a region within which deposit heights exceed
        $0.1\textrm{m}$.  (Both these regions are localised near to the source
        flux.) Dashed black lines show the contours of the corresponding
        analytical similarity solution, given in Eqs.~\eqref{eq:sim soln 1}
        and~\eqref{eq:sim soln 2}, at $0.1$, $0.2$ and $0.3\mathrm{m}$.  The
        grid resolutions are $\Delta x =$ (a)~$1\textrm{m}$ and
        (b)~$0.5\textrm{m}$.
        }%
    \label{fig:const slope mild}%
    \end{centering}%
\end{figure}
Figure~\ref{fig:const slope mild} shows contour plots of the flow depth
after $t = 800\textrm{s}$ for two different initial
slope angles, (\emph{a})~$\vartheta = 1^\circ$ and 
(\emph{b})~$\vartheta = 5^\circ$.
Both results feature 
significant erosion of the bed: as much as (\emph{a})~$-0.62\textrm{m}$
and (\emph{b})~$-2.5\textrm{m}$ respectively.
However, this is confined to compact regions near the source.
Blue contours bound regions within which the net change in
bed height due to erosion is below $-0.1\textrm{m}$.
The $5^\circ$ flow features a region within which net deposition of the excavated
bed material exceeds $0.1\textrm{m}$, indicated by the dashed red bounding curve.
Outside these regions, the down- and cross-slope expansion of the flow is
approximately steady and the bed exchange term attains a dynamic balance.
Indeed, across the extent of the flow that is inundated up to a depth of
$0.1\textrm{m}$ and more than $200\textrm{m}$ from the source, we
compute $\max(|E - D|/E)\leq$ (\emph{a})~$6\times 10^{-3}$ and
(\emph{b})~$0.04$.

\begin{figure}
    \begin{centering}
    \includegraphics[width=\textwidth]{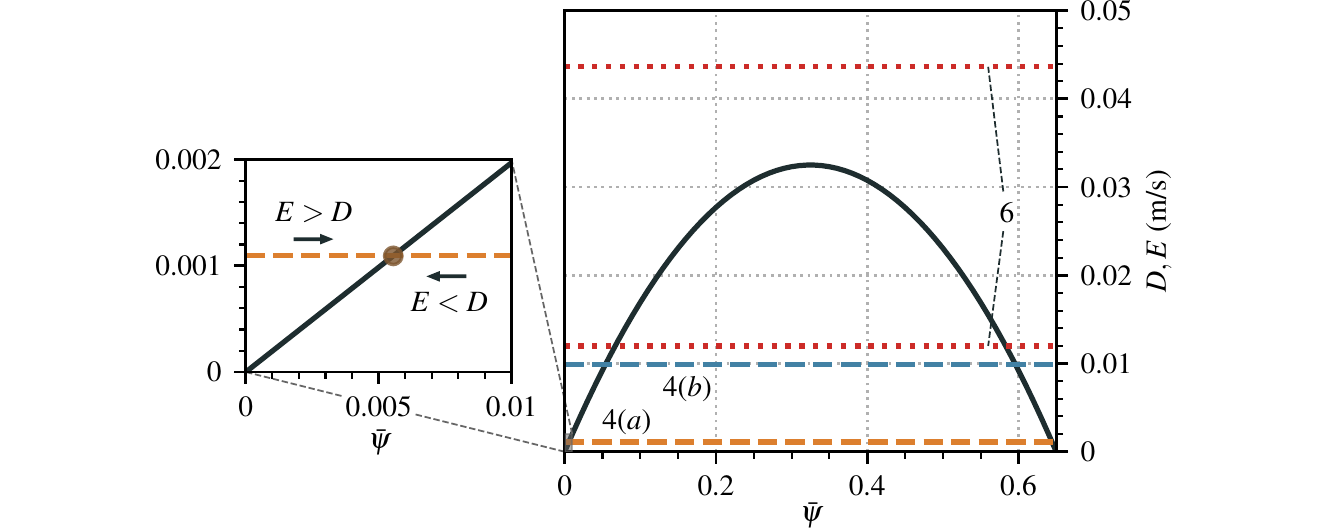}%
        \caption{Erosion and deposition rates from the constant slope
        simulations as a function of flow solid fraction $\bar{\psi}$. The solid
        black curve is the deposition rate $D(\bar{\psi})$, while the orange and
        blue dashed curves show erosion rate $E(|\bar{\vect{u}}_{\max}|)$, where
        $|\bar{\vect{u}}_{\max}|$ is the maximum flow speed within the inundated
        area of the flows presented in Figs.~\ref{fig:const slope
        mild}(\emph{a}) and~(\emph{b}), as labelled. The dotted red curves show
        $E$ for $|\bar{\vect{u}}| = 9.69\textrm{m}/\textrm{s}$ and
        $5.08\textrm{m}/\textrm{s}$. These values are the maximum flow speeds
        (to 3~s.~f.) computed for the result given below in Fig.~\ref{fig:const
        slope severe 1}, over its two distinct regimes, which we designate as $x
        < 250\textrm{m}$ and $x > 250\textrm{m}$ respectively.}%
        \label{fig:EandD}%
    \end{centering}
\end{figure}

The equilibration of erosion and deposition occurs because, for each speed
$|\bar{\vect{u}}|$
present in the flow, there exists a stable sediment
concentration that the system relaxes onto at each point.
The essential mechanism for this was explored previously in
Ref.~\cite{Langham2021}. We illustrate it
graphically in Fig.~\ref{fig:EandD}. 
This shows the deposition rate curve and example erosion rate curves, each as a function of
$\bar\psi$.  For both the $1^\circ$ and $5^\circ$ flows, these intersect at two
locations where $D$ and $E$ are exactly balanced.  As highlighted in the
enlarged portion of Fig.~\ref{fig:EandD}, perturbations away from the more dilute
intersection point induce negative feedback from the morphodynamics that acts to
return the system to equilibrium.  While the flow speed (and hence the erosion
rate) varies spatially throughout each flow, it nowhere reaches a value that
would cause $E$ to exceed the maximum deposition rate, $D(\psi_b/2)$.
Therefore, at each inundated point in these solutions, a stable equilibrium
exists and we have verified from the data that these equilibria are
approximately attained outside the vicinity of the source flux region.

Since these flows are approximately steady with almost vanishing morphodynamics
downstream from the source, it is possible to construct analytical solutions by
hypothesising that these properties become asymptotically exact at long times.
In Appendix~\ref{appendix:similarity}, we
show how to derive these for a general class of drag closures.
When specialised to the current case of Ch\'ezy drag, 
we recover the following solutions, with a form
previously obtained by Bonnecaze \& Lister~\cite{Bonnecaze1999}:
\begin{subequations}
\begin{equation}
    H(x,\xi) = C_H x^{-1/4} (1 - \xi^2), 
    \quad w_p(x) = C_w x^{3/8},
    \quad \bar{u}(x,\xi) = \Lambda H^{1/2}, 
    \label{eq:sim soln 1}%
    \tag{\theequation\emph{a--c}}%
\end{equation}
\end{subequations}
where $\xi = y / w_p(x)$, with $w_p(x)$ 
denoting the width of the inundated region
and $C_H$, $C_w$, $\Lambda$ are constants given by 
\begin{subequations}
\begin{equation}
    C_H = \left(
    \frac{4Q^2 \sin\vartheta}{3\pi \Lambda^2}
    \right)^{\!\!1/4}
    \!\!,\quad
    C_w = \left(
    \frac{16 C_H}{3\sin\vartheta}
    \right)^{\!\!1/2}
    \!\!,\quad
    \Lambda = \left(g \sin\vartheta\cos^2\!\vartheta\right)^{\!\!1/2}
    \!\!.
    \label{eq:sim soln 2}%
    \tag{\theequation\emph{a--c}}%
\end{equation}
\end{subequations}
Contours of these asymptotic solutions are included in Fig.~\ref{fig:const slope
mild}, plotted with dashed black lines. In both cases, in the interior of the
solution these align almost
perfectly with the corresponding contours of the simulated depth field,
validating the numerical results.

\subsubsection{Steeper slopes.}
\begin{figure}
    \begin{centering}%
    \includegraphics[width=\textwidth]{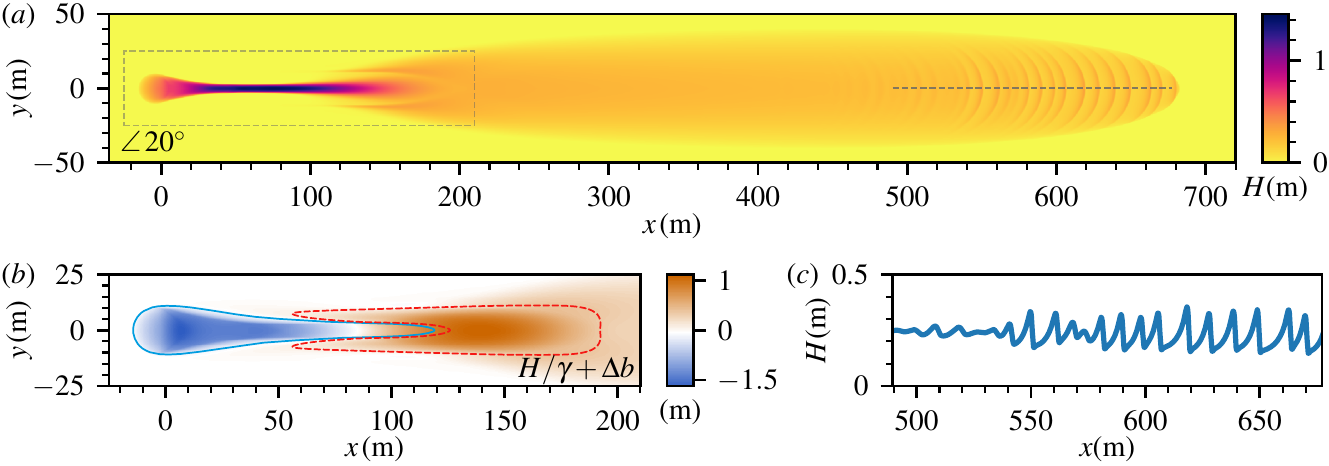}%
        \caption{%
        Flow propagating on an incline with a constant initial slope of
        $20^\circ$, fed by a
        $50\mathrm{m}^3/\mathrm{s}$ point source flux at the origin, at
        $t=160\textrm{s}$.
        The numerical resolution is $\Delta x = 0.25\textrm{m}$.
        (\emph{a})~Colour map of flow depth. 
        (\emph{b})~Sum of the flow depth, projected vertically, $H/\gamma$ 
        and the vertical change in bed height $\Delta b
        = b|_{t=160\textrm{s}} - b|_{t=0\textrm{s}}$, within the
        self-channelised region [outlined by the dashed grey rectangle
        in~(\emph{a})]. Points in the flow that lie below (above) the initial bed
        height are coloured blue (orange). The colour scale is linear, either
        side of zero. The solid blue and red dashed contours bound regions that
        are net erosive and net depositional: they show $\Delta b =
        -0.1\textrm{m}$ and $0.1\textrm{m}$ respectively.  The range of change
        in bed height over this area is between $-1.7\textrm{m}$ and
        $+1.1\textrm{m}$ (to $2$~s.\ f.).
        (\emph{c})~Flow depth along the centreline interval indicated by the grey
        dashed line in~(\emph{a}).
        }
    \label{fig:const slope severe 1}%
    \end{centering}%
\end{figure}

At steeper slope angles, there are richer interactions between the flow and
morphodynamics.  In Fig.~\ref{fig:const slope severe 1}(\emph{a}), we plot
contours of flow depth for a flow on an initially constant $20^\circ$ slope
after~$160\textrm{s}$.
We observe the flow splitting into two distinct regimes.  Immediately downstream
of the source, the flow is confined within a channel that it has excavated from
the bed, extending for roughly~$100\textrm{m}$ (at the time of the plotted
snapshot).  Detail of the morphodynamics in this region is given in
Fig.~\ref{fig:const slope severe 1}(\emph{b}). Points coloured blue in this plot
lie below the original bed elevation and thus indicate flow within deeply eroded
regions, whose lateral slopes constrain the material.  This 'self-channelised'
region grows deeper and longer in time, as more of the bed is removed. Further
downstream, there is a region of deposit formed by some of the eroded sediment,
followed by a longer, unconfined spreading flow.  The flow front is travelling
faster than the downstream expansion of the self-channelised region, which
explains why most of the flow is unconfined at this point.  The downstream flow
does not erode or deposit a significant amount of sediment, indicating that it
is in approximate morphodynamic equilibrium. However, because it is propagating
on a steep slope, the flow is nonetheless vulnerable to the classical roll wave
instability~\cite{Jeffreys1925,Cornish1934,Dressler1949,Needham1984}, the
emergence of which can be seen around $x \gtrsim 500\textrm{m}$. A
one-dimensional cross-section of the free surface waves that the flow generates
is plotted in Fig.~\ref{fig:const slope severe 1}(\emph{c}).

The difference in behaviour between the two regimes may be understood by
referring back to Fig.~\ref{fig:EandD}. Its dotted red curves show the maximum
erosion rates for the flow inside and outside the channelised region [taken to
be the area of Fig.~\ref{fig:const slope severe 1}(\emph{b})].  Flow within the
channel reaches high speeds that cause the erosion rate to exceed the maximum
deposition rate.  This leads to sustained erosion and the consequent
accumulation of sediment concentration in the eroding region, which is advected
downstream. As the sediment propagates away from the source, its concentration
diminishes and flow speeds are reduced, so eventually $D > E$, leading to the
formation of a deposit. The flow that spreads further downstream reaches a
dilute morphodynamic equilibrium, analogous to what is observed in the $1^\circ$
and $5^\circ$ slopes (see Fig.~\ref{fig:const slope mild}).

\begin{figure}
    \begin{centering}%
    \includegraphics[width=\textwidth]{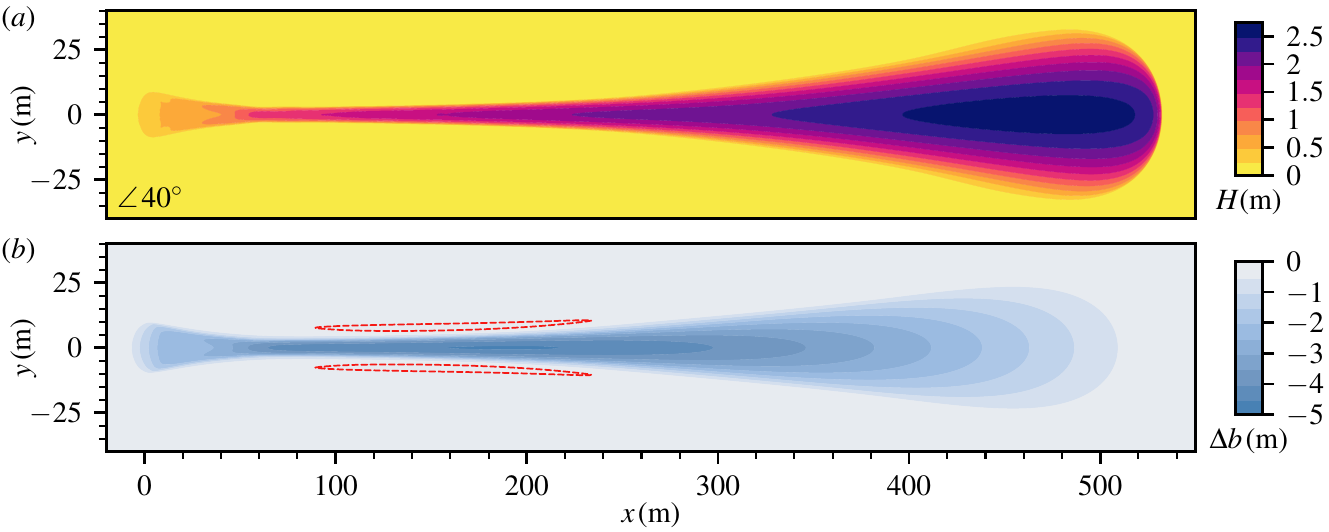}%
        \caption{A 'super-erosive' flow state at $t = 80\textrm{s}$. 
        The initial bed is a constant slope at
        $40^\circ$ inclination, fed by a
        $50\mathrm{m}^3/\mathrm{s}$ point source flux at the origin.
        The grid resolution is $\Delta x = 0.25\textrm{m}$.
        (\emph{a})~Contours of flow depth.
        (\emph{b})~Bed elevation change $\Delta b = b|_{t=80\textrm{s}} -
        b_{t=0\textrm{s}}$. 
        Blue filled contours indicate regions of net erosion.
        Dashed red contours are regions within which net deposition exceeds
        $+0.1\textrm{m}$. The maximum deposition in these regions is
        $+0.22\textrm{m}$~(2~s.~f.).}
    \label{fig:const slope severe 2}%
    \end{centering}%
\end{figure}

At yet higher slope angles, a new state emerges, in which erosion occurs so
rapidly that the expansion of eroded region matches the flow front velocity.  In
Fig.~\ref{fig:const slope severe 2}(\emph{a,b}), we plot contours of the flow
depth and the change in bed height, for a solution at initial slope
angle $40^\circ$, after $80\textrm{s}$.
The striking teardrop shape of the inundated area is almost entirely contained
within a region that is net erosive.  Indeed, so much material is entrained that
the flow attains the maximum sediment concentration $\bar\psi = \psi_b$
throughout this region. Due to the hindered settling law, this renders
deposition essentially nonexistent, save for some narrow bands at the edges of
the flow.  While the flow is initially driven by the source flux, its volume
quickly becomes dominated by entrained material.  In fact, once established, the
state is self-perpetuating and continues to propagate even if the source is
turned off.  This is possible due to an erosive `domino effect'---rapid erosion
liberates gravitational potential energy from the bed, which accelerates the
front and increases the erosion rate in turn.  The principal restoring force
available, basal drag, is unable to match the accumulated forcing from the
slope-parallel weight, which grows more rapidly as the flow progresses.

\subsection{Flow between two uniform slopes.}
We conclude this section with an example of flow that propagates from a steep
slope to a shallower one, as typically occurs when debris flows are initiated on
mountainsides.
Specifically, we consider the following initial surface:
\begin{equation}
    b(x,y,0) = -\frac{(s_1+s_2)x}{2}
    + \frac{(s_1 -
    s_2)\lambda}{2}\log\left[\cosh\left(\frac{x}{\lambda}\right)\right]\!\!.
\end{equation}
This constructs a slope that transitions between gradients $s_1$ upstream
($x\to-\infty$) and $s_2$ downstream ($x\to\infty$), over a characteristic
length scale $\lambda$.  

\begin{figure}
    \begin{centering}%
    \includegraphics[width=\textwidth]{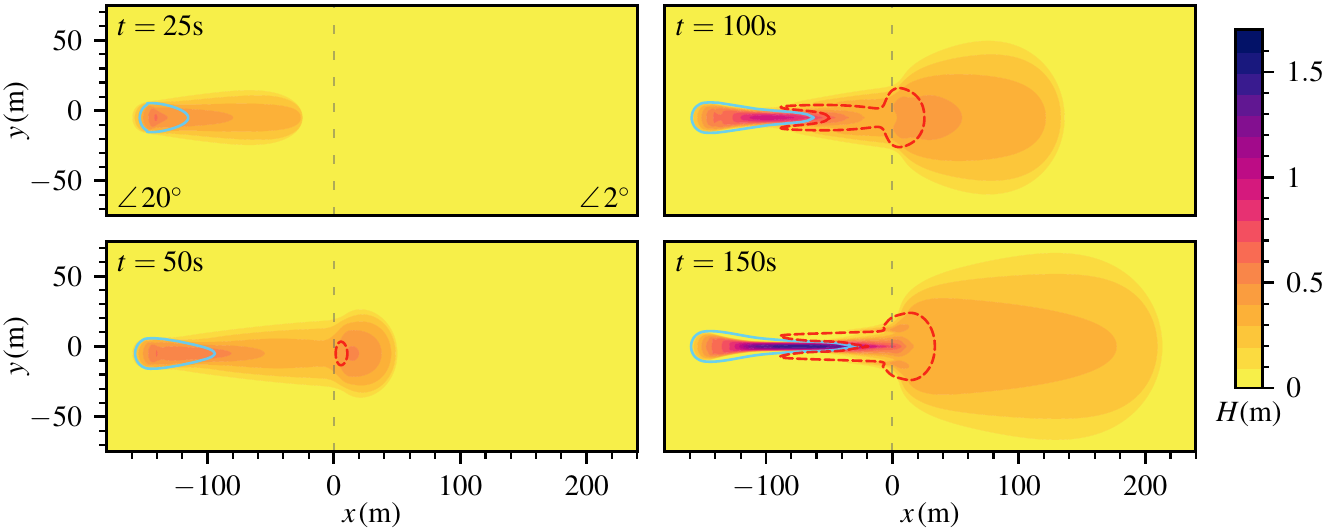}
        \caption{Simulated flow on an initially constant $20^\circ$ slope, that
        smoothly transitions to an initially constant $2^\circ$ slope
        downstream. The centreline of this transition, at $x=0\textrm{m}$, is
        shown in dashed grey.  The flow is fed by a
        $50\textrm{m}^3/\textrm{s}$ point source flux at $(x,y) =
        (-150\textrm{m},0\textrm{m})$.  Four filled contour maps of flow depth
        are plotted, at times $t = 25\textrm{s}$, $50\textrm{s}$,
        $100\textrm{s}$ and $150\textrm{s}$, as indicated.  Solid blue contours
        enclose regions where erosion has reduced the bed elevation by more than
        $0.1\textrm{m}$ and red dashed contours indicate regions where net
        deposits exceed $0.1\textrm{m}$.  The grid resolution is $\Delta x =
        0.25\textrm{m}$.
        }
    \label{fig:bislope}%
    \end{centering}%
\end{figure}

In Fig.~\ref{fig:bislope}, we present four progressive snapshots of a simulation
with $s_1 = \tan(20^\circ)$, $s_2 = \tan(2^\circ)$ and $\lambda = 5\textrm{m}$,
initiated $150\textrm{m}$ upstream from the change in slope.
The flow shares properties observed in both the steep and gentle
constant slope simulations of the previous subsection.  In the first snapshot,
at $t = 25\textrm{s}$, the flow has progressed rapidly.  It has eroded some of
the bed near the source, but not enough to confine the bulk of the current.  At
$t = 50\textrm{s}$, the flow has reached the milder slope, slowing the speed of
the front and causing it to expand laterally.  When the flow decelerates, this
the reduces erosion rate, which in turn reduces the sediment carrying capacity
of the flow.  This leads to the creation of a deposit formed from some of the
flow material excavated upstream.  At $t = 100\textrm{s}$, the upstream
erosional region is now deep enough for the flow to self-channelise, as observed
above in the $20^\circ$ simulation presented in Fig.~\ref{fig:const slope severe
1} [see especially panel~(\emph{b})].  As the simulation progresses to $t =
150\textrm{s}$, the erosional region deepens and gradually invades its own
deposit.  Far downstream, the flow has matured into a steady spreading
equilibrium, of the kind observed in Fig.~\ref{fig:const slope mild}.

\section{Discussion}
\label{sec:discussion}%
In this paper, new depth-averaged governing equations were derived for
morphodynamic shallow flows with a suspended sediment phase, which are valid for
weakly curved topographies that feature an arbitrary range of basal gradients.
This advance is crucial for future modelling studies in geophysical settings,
where flows transition from very steep slopes onto gentler ones, such as debris
flows initiated in mountainous regions~\cite{Jakob2005} and volcanic
flows~\cite{Pierson1990,Scott2005}.  Our equations are formulated from a general
perspective that allows them to be specialised to a variety of settings, from
dilute suspensions to mudflows and granular media, by specifying appropriate
modelling closures.  Moreover, they are far simpler than shallow models that
incorporate the full effects of arbitrary surface curvature making them more
tractable to simulate from the perspective of practical natural hazards
assessment.

To this end, we developed and implemented a strategy for numerically solving the
governing equations using an operator splitting approach that enables existing
non-morphodynamic finite volume codes to be extended to include coupled bed
evolution and a sediment phase.  Desirable properties of the underlying shallow
flow scheme may be retained this way---in our implementation (which extends
methods reported in Ref.~\cite{Chertock2015a}) this includes exact numerical
integration of steady states (well balancing), though some new steps in the
scheme were necessary to accommodate this for morphodynamic flows, as detailed
in Sec.~\ref{sec:numerics}.  In this way, there is the potential to extend the
operator-split morphodynamic scheme, as and when technical advances in shallow
water methods become available. Though it might also be anticipated that mass
conservation should follow automatically (given that finite volume schemes are
typically conservative by construction), we have had to resolve multiple
subtleties involving the geometric transformation of volume elements and
book-keeping of morphodynamic transfers between the bed and flow. Failure to
take similar precautions may result in schemes that are systematically biased
towards gaining or losing mass in context-dependent ways.

Our numerical implementation incorporates terms corresponding to turbulent
momentum diffusivity, given in Eq.~\eqref{eq:diffusion}.  Though often neglected
in these models (Ref.~\cite{Simpson2006} is a rare exception), this physics
offers a way to ensure that the governing equations are well posed when
integrated as an initial-value problem~\cite{Langham2021}.  Our demonstration of
the problem in Fig.~\ref{fig:ill posedness} highlights the issue that faces
unregularised schemes: finer discretisation leads to divergence of the
numerically computed flows, rather than convergence. The severe instability that
underpins the phenomenon only emerges at length scales shorter than grid
resolutions typically used in practical simulations, which suggests a possible
reason why the issue may have previously gone unnoticed in susceptible models.
This highlights the necessity of robust convergence testing for morphodynamic
models, especially when they may be used in practical hazards assessment and
other sensitive applications.  In addition to eddy diffusivity, the issue may be
alleviated by adding bed load fluxes to the morphodynamic equation (which some
models already feature~\cite{Wu2007,Murillo2010,Benkhaldoun2013,Liu2017}),
though we note that well posedness is contingent on the choice of flux closure
in this case~\cite{Chavarrias2018,Langham2021}.

The results presented in Sec.~\ref{sec:results} as demonstrations of the model
in operation merit some discussion in their own right. Though the closures we
have adopted are somewhat simplified, they are nonetheless sufficient to offer
some phenomenological insight into the basic interactions of shallow flows with
the processes of erosion and deposition.  Our survey of flows spreading from
point flux sources on initially constant slopes reveals some interesting flow
regimes.  On mild slopes, basal drag is sufficiently weak that sediment
entrainment adopts a dynamic balance with the deposition rate throughout the flow bulk,
thereby nullifying the net effect of morphodynamics. We demonstrated that this
is the consequence of an attracting fixed point in the bed evolution equation
that always exists when erosion lies below a certain threshold.  Within this
regime, it is possible to construct similarity solutions for the flow that reach
very good agreement with the numerical simulations.  In
Appendix~\ref{appendix:similarity}, we derived formulae for these solutions that
encompass a broad class of basal drag closures.  These may prove valuable (as
they have done here) for testing numerical schemes, since analytical solutions
are usually difficult to obtain for morphodynamic flows.

On steeper slopes, which drive higher flow speeds, the entrainment rate is able
to exceed the maximum rate of sediment deposition, leading to substantial
erosion. The bifurcation to this state can be quite dramatic, since sediment
uptake enhances the local solids concentration, which reduces the deposition
rate via the effect of hindered settling and thereby increases net erosion in
turn. In this regime, spreading flows cut grooves in the bed that are
preferentially focussed
along their centreline (where flow speed is maximised, in these simple states).
These grooves ultimately grow into deep channels, leading to flow self-confinement.
In a related, but more extreme regime observed at higher slope angles, the flow
develops into a highly concentrated self-accelerating front, due to a positive
feedback loop between gravitational forcing and entrainment, which outscales the
restorative influence of drag.  
To our knowledge, such a phenomenon has not previously been identified in debris
flows. However, it shares clear parallels with observations of self-accelerating
turbidity currents, whose entrainment of basal sediment has been ascribed as a
direct mechanism for their
sustenance~\cite{Parker1982,Parker1986,Blanchette2005}. In our simulations, the
self-accelerating states grow rapidly and without bound.  In natural
applications, this growth will instead be limited by the finite extent of steep
slopes and the availability of readily erodible material, which is bounded
because the subsurface typically consists of an erodible layer of soils and
unconsolidated sediment underlain by much less erodible
bedrock~\cite{Heimsath2012}.  Nevertheless, it could be that these processes are
transiently responsible for observations of particularly intense erosion from
debris flows~\cite{Jakob2005,Stock2006,Shen2020}.  

While we would anticipate some quantitative changes in our results if our
simplified closures were substituted with empirical formulae fitted to real
scenarios, we expect the qualitative phenomena described here to be robust.
Relatedly, we concluded our results with a brief example flow propagating from a
steeper to a shallower slope.  We note that most of its qualitative features
could have been anticipated in advance, by referring to flow regimes already
identified in the constant slope cases. This suggests that insights gleaned from
simple morphodynamic flows may be useful for understanding dynamics on more
complex terrains.

\vskip6pt

\enlargethispage{20pt}



\aucontribute{AJH and JL derived the model and the mathematical results. MJW and
JL co-wrote the simulation code and developed the numerical scheme. LTJ and JCP
helped to test the code and contributed to research discussions throughout
development along with the other authors. JL conducted the simulations, wrote
the paper and all authors proof-read the final manuscript.}

\competing{We declare that we have no competing interests.}

\funding{
Much of the work for this article was conducted as part of the Newton Fund grant
NE/S00274X/1. Additionally, we have benefited from various other
sources during the course of the study:
JL acknowledges funding from the EPSRC Impact Acceleration Account
EP/X525674/1; MJW acknowledges a NERC
Knowledge Exchange Fellowship NE/R003890/1;
AJH acknowledges the Royal Society grant APX/R1/180148;
LTJ and JCP acknowledge funding from the
UKRI Global Challenges Research Fund grant NE/S009000/1 and
JCP acknowledges a University of Bristol Research Fellowship.
}

\ack{We are grateful for discussions with Chris Johnson at an early stage of
development and to our many geological colleagues in IG-EPN (Ecuador), INGEMMET
(Per\'{u}) and PHIVOLCS (Philippines), who helped to test versions of the
numerical code.}


\appendix

\section{Lake at rest}
\label{appendix:lar}%
Here, we show that time derivatives computed by our numerical method are exactly
zero for `lake-at-rest' initial conditions, which satisfy Eq.~\eqref{eq:lake at
rest analytic}. 
We denote by $\eta_c$ and $\bar\rho_c$, the spatially-constant
values of $\eta$ and $\bar\rho$ respectively.
As discussed above, the only nontrivial cases to consider are the
momentum equations, for which we must verify that the discretised forms of the
hydrostatic pressure gradient must exactly balance.
These are given in Eqs.~\eqref{eq:hydrostatic pressure discrete} and~\eqref{eq:grav forcing
discrete}. Therefore, for the $\bar u$-momentum equation, we must verify that
\begin{equation}
    \left[1 + (b_y)_{i,j}^2\right]
    \frac{\partial \zeta}{\partial x}(x_i,y_j)
    - (b_x)_{i,j} (b_y)_{i,j} \frac{\partial \zeta}{\partial y}(x_i,y_j)
    =
    -\bar\rho_c g (\eta_c - b_{i,j}) (b_x)_{i,j},
    \label{eq:lake at rest balance}%
\end{equation}
where $\zeta(x_i,y_j) = \frac{1}{2}\bar\rho_c g (\eta_c - b_{i,j})^2$.
Using Eqs.~\eqref{eq:b ifaces} and~\eqref{eq:b centre}, we
compute
\begin{subequations}
\begin{align}
    \frac{\partial \zeta}{\partial x}(x_i,y_j)
    &=
    \frac{\bar\rho_c g}{2\Delta x}
    \left[
        (\eta_c - b_{i+1/2,j})^2 - (\eta_c - b_{i-1/2,j})^2
        \right]\\
    &= -\bar\rho_c g\left[
        (2\eta_c - b_{i-1/2,j} - b_{i+1/2,j})\frac{(b_{i+1/2,j}-b_{i-1/2,j})}{2\Delta
        x}
        \right]\\
    &= -\bar\rho_c g (\eta_c - b_{i,j})(b_x)_{i,j}.
\end{align}
\end{subequations}
Therefore, we must show that the remaining terms on the left-hand side of
Eq.~\eqref{eq:lake at rest balance} sum to zero.
This is clear after computing
\begin{subequations}
\begin{align}
    (b_y)_{i,j} \frac{\partial \zeta}{\partial x}(x_i, y_j)
    &=
    \frac{\bar\rho_c g}{2\Delta x \Delta y}
    (b_{i,j+1/2}-b_{i,j-1/2})
    \left[
        (\eta_c - b_{i+1/2,j})^2 - (\eta_c - b_{i-1/2,j})^2
        \right]\label{eq:lar step 2}\\
    &= 
    \frac{\bar\rho_c g}{2\Delta x \Delta y}
    (b_{i+1/2,j} - b_{i-1/2,j})
    \left[
        (\eta_c - b_{i,j+1/2})^2 - (\eta_c - b_{i,j-1/2})^2
        \right]\label{eq:lar step 3}\\
    &= (b_x)_{i,j}\frac{\partial \zeta}{\partial y}(x_i,y_j),
\end{align}
\end{subequations}
and multiplying both sides by $(b_y)_{i,j}$.  To swap the indices between
Eqs.~\eqref{eq:lar step 2} and~\eqref{eq:lar step 3}, we rearranged the terms
and used the fact that $b_{i+1/2,j}+b_{i-1/2,j} = b_{i,j+1/2} + b_{i,j-1/2}$.
The balances for the $\bar v$-momentum equation are obtained by the same
reasoning due to symmetry.

\section{Similarity solutions for mild constant slopes}
\label{appendix:similarity}%
If the morphodynamics processes are in equilibrium, then $\morpho = 0$ and the
bulk density $\bar\rho$ is constant. 
Since motion is primarily downslope, we expect $\bar{u} \gg \bar{v}$ and
therefore, $|\bar{\vect{u}}| = \bar{u}/\cos(\vartheta)$ to leading order. 
Furthermore, we seek solutions for steady flow, whose spatial variation is
sufficiently small that the effects of eddy viscosity may be neglected.
These assumptions lead to considerable simplifications of the governing
equations~\eqref{eq:dqdt partial}--\eqref{eq:S2}.
The dominant
balances in the momentum equations become
\begin{subequations}
\begin{equation}
    gH \sin(\vartheta) = \frac{\tau_b}{\bar\rho}, \quad
    \textrm{and}\quad
    gH \frac{\partial H}{\partial y} =
    -\frac{\tau_b}{\bar\rho}\frac{\bar{v}}{\bar{u}}.
    \tag{\theequation\emph{a,b}}%
\end{equation}
    \label{eq:steady balances}%
\end{subequations}
For many drag formulations, Eq.~(\ref{eq:steady balances}\emph{a}) may be
rearranged to give $\bar{u} = \Lambda H^m$, where $\Lambda$ and $m$ are
constants particular to the choice of $\tau_b$~(e.g.\ see \cite{Langham2022}).
This includes the Ch\'ezy drag closure employed above [Eq.~\eqref{eq:chezy}],
for which $m=1/2$. Substituting the general expression into the mass
conservation equation gives
\begin{equation}
    \frac{\partial~}{\partial x}(H^{m+1}) =
    \frac{1}{\sin(\vartheta)}\frac{\partial~}{\partial y}
    \left(
    H^{m+1} \frac{\partial H}{\partial y}
    \right)\!.
    \label{eq:mass conserv sim}%
\end{equation}
This is subject to a flux condition
\begin{equation}
    \int_{-w_p(x)}^{w_p(x)} H^{m+1}\, \mathrm{d}y = \frac{Q}{\Lambda},
    \label{eq:flux condition}%
\end{equation}
for all $x > 0$ (i.e.\ downstream of the source), where $w_p(x)$ denotes the distance from the symmetry line
$y=0$ to the perimeter of the wetted region in the $y$-direction.

Since there are no external length scales present in the system, we expect the
asymptotic solutions to be self-similar.  The essential scalings
present in Eqs.~\eqref{eq:mass conserv sim} and~\eqref{eq:flux condition} are
$H/w_p^2 \sim \sin(\vartheta)/x$ and $H^{m+1} \sim Q/(\Lambda w_p)$.
These motivate the following ansatz
\begin{subequations}
\begin{equation}
    H(x,\xi) =
    C^2\left[\frac{Q^2}{\Lambda^2}\frac{\sin(\vartheta)}{x}\right]^{\!\frac{1}{2m+3}}\!\!F(\xi),
    \quad w_p(x) = C\left[\frac{Q}{\Lambda}
    \left(\frac{x}{\sin(\vartheta)}\right)^{\!m+1}
    \right]^{\frac{1}{2m+3}}\!,
    \tag{\theequation\emph{a,b}}%
\end{equation}
\end{subequations}
where $\xi = y / w(x)$ and $C$ is a constant to be determined.  On substituting these
into Eq.~\eqref{eq:mass conserv sim}, we arrive at an ordinary differential
equation for the cross-stream dependence,
\begin{equation}
    -\frac{m+1}{2m+3}\frac{\mathrm{d}~}{\mathrm{d}\xi}\left[\xi F(\xi)^{m+1}\right]
    = \frac{\mathrm{d}~}{\mathrm{d}\xi}\left[F(\xi)^{m+1}
    \frac{\mathrm{d}F}{\mathrm{d}\xi}\right].
\end{equation}
By integrating and applying the boundary condition $F(\pm 1) = 0$, we find that
$F(\xi) = F_0 (1-\xi^2)$, where $F_0 = \frac{m+1}{2(2m+3)}$.
Then, by evaluating Eq.~\eqref{eq:flux condition}, 
$C$ may be determined to complete the solutions, which are given compactly as
\begin{subequations}
\begin{equation}
    H(x,\xi) = \left[
        \frac{F_0 Q^2}{\Upsilon^2\Lambda^2} \frac{\sin(\vartheta)}{x}
    \right]^{\!\frac{1}{2m+3}}(1-\xi^2),
    \quad
    w(x) = \left[
        \frac{Q}{\Upsilon \Lambda}
        \left(
        \frac{x}{F_0\sin(\vartheta)}
        \right)^{\!m+1}
        \right]^{\frac{1}{2m+3}}\!\!,
    \tag{\theequation\emph{a,b}}%
\end{equation}
\end{subequations}
where $\Upsilon = \int_0^\pi \sin^{2m+3}(\phi)
\,\mathrm{d}\phi$.  To specialise these to a particular drag law, one simply
specifies $m$.

\bibliographystyle{RS} 
\bibliography{LaharBib} 

\end{document}